\documentclass{aa} 
\usepackage{txfonts}
\usepackage[para,online,flushleft]{threeparttable}
\usepackage{subcaption}
\usepackage{hyperref}
\DeclareUnicodeCharacter{2212}{-}

\begin{document}

    \title{FORECAST: A flexible software to forward model cosmological hydrodynamical simulations mimicking real observations}
    \titlerunning{FORECAST: A tool to forward model cosmological hydrodynamical simulations}
    
    \author{Flaminia Fortuni \inst{1}
        \and Emiliano Merlin \inst{1}
        \and Adriano Fontana \inst{1}
        \and Carlo Giocoli \inst{3,4} 
        \and Erik Romelli \inst{2}
        \and Luca Graziani \inst{5}
        \and Paola Santini \inst{1}
        \and Marco Castellano \inst{1}
        \and Stéphane Charlot \inst{6}
        \and Jacopo Chevallard \inst{7}} 

    \institute{INAF - Osservatorio Astronomico di Roma, via Frascati 33, 00078 Monte Porzio Catone (Roma), Italy
        \email{flaminia.fortuni@inaf.it}
        \and INAF - Osservatorio Astronomico di Trieste, Via Tiepolo 11, I-34131 Trieste, Italy
        \and INFN - Sezione di Bologna, Viale Berti Pichat 6/2, 40127 Bologna, Italy
        \and INAF - Osservatorio di Astrofisica e Scienza dello Spazio di Bologna, via Gobetti 93/3, I-40129 Bologna, Italy
        \and Dipartimento di Fisica, Università di Roma "La Sapienza", Piazzale Aldo Moro 5, I-00185 Roma, Italy
        \and Sorbonne Université, CNRS, UMR 7095, Institut d’Astrophysique de Paris, 98 bis bd Arago, 75014 Paris, France
        \and Department of Physics, University of Oxford, Denys Wilkinson Building, Keble Road, Oxford OX1 3RH, UK}
    
    \date{Received date /   Accepted date } 
    
    \abstract {Comparing theoretical predictions to real data is crucial to properly formulate galaxy formation theories. However, this is usually done naively considering the direct output of simulations and quantities inferred from observations, which can lead to severe inconsistencies.}
    {We present FORECAST, a new flexible and adaptable software package that performs forward modeling of the output of any cosmological hydrodynamical simulations to create a wide range of realistic synthetic astronomical images, and thus providing a robust foundation for accurate comparison with observational data. With customizable options for filters, field-of-view size, and survey parameters, it allows users to tailor the synthetic images to their specific requirements.}
    {FORECAST constructs a light cone centered on the observer's position exploiting the output snapshots of a simulation and computes the observed flux of each simulated stellar element, modeled as a single stellar population, in any chosen set of passband filters, including $k$ correction, intergalactic medium absorption, and dust attenuation. These fluxes are then used to create an image on a grid of pixels, to which observational features such as background noise and PSF blurring can be added. This allows simulated galaxies to be obtained with realistic morphologies and star formation histories.} 
    {As a first application, we present a set of images obtained exploiting the \textsc{IllustrisTNG} simulation, emulating the GOODS-South field as observed for the CANDELS survey. We produced images of $\sim$200 sq. arcmin., in 13 bands (eight \textit{Hubble Space Telescope} optical and near-infrared bands from ACS $B$435 to WFC3 $H$160, the \textit{VLT} HAWK-I $Ks$ band, and the four IRAC filters from \textit{Spitzer}), with depths consistent with the real data. 
    We analyzed the images with the same processing pipeline adopted for real data in CANDELS and ASTRODEEP publications, and we compared the results against both the input data used to create the images and the real data, generally finding good agreement with both, with some interesting exceptions which we discuss. As part of this work, we have released the FORECAST code and two datasets. The first is the CANDELS dataset analyzed in this study, and the second dataset emulates the \textit{JWST} CEERS survey images in ten filters (eight NIRCam and two MIRI) in a field of view of 200 sq. arcmin. between $z$=0-20.}
    {FORECAST is a flexible tool: it creates images that can then be processed and analyzed using standard photometric algorithms, allowing for a consistent comparison among observations and models, and for a direct estimation of the biases introduced by such techniques.}%

    
    \maketitle
    
    \section{Introduction}

    In the past two decades, several imaging and spectroscopic surveys have revolutionized our understanding of galaxies across the electromagnetic spectrum (e.g., \citealt{colless03,abazajian03,giav04,lilly07,scoville07,driver09,grogin11,koek11,brammer12,tom14,pente18}). These surveys have provided observations of thousands of galaxies, enabling their systematic study and classification at different epochs. Advancements in technological capabilities are pushing the boundaries of space exploration, allowing us to observe the Universe farther in space and further in time, reaching the dawn of the first lights. \textit{JWST} is providing exquisite data on the early stages of galaxy evolution, yielding unprecedented results that are challenging our understanding of galaxy formation and evolution (e.g.,  \citealt{morishita22,treu22,r-b22,castellano22,naidu22,finkelstein22a,curtis-lake22,robertson22,yan23,donnan23,harikane23}). The future ahead holds even greater excitement as it introduces a new generation of telescopes, including \textit{Euclid}, the \textit{Nancy Grace Roman Telescope}, the \textit{European-Extremely Large Telescope}, and the \textit{Vera Rubin Observatory}.
    
    High-quality data must be compared to precise theoretical predictions. Cosmological simulations, encompassing a wide range of approaches, have achieved a remarkable level of sophistication, producing detailed characterization of the Universe across an extensive range of spatial and temporal scales. Hydrodynamical simulations, in particular, self-consistently simulate the evolution of both dark matter and baryons, providing insights into the complex nonlinear processes involved in the growth of cosmic structures, including the formation of galaxies, the interplay between gas dynamics and gravitational forces, and the emergence of large-scale cosmic filaments.
    Their successful reproduction of observable properties and scaling relations of real galaxies (e.g., \citealt{hern96,choi10,devr10,park12,genel14,pill18,kavi17,nelson18,vogel18,cui21,dicesare23}) establishes them as effective guidance for interpreting observational data.

    However, comparing simulation predictions to observed data requires establishing a coherent linkage between the physical and the observable domains. This can be achieved (i) by moving from the observational to the physical quantities, using the features of real imaging data to estimate a set of underlying physical parameters or models (indirect approach), or (ii) by going in the opposite direction, reproducing and mimicking observations from theory (forward approach). Converting photometric or spectroscopic data into physical quantities using the indirect approach is a common practice in astronomical research. When only imaging data are available, which is often the case for large-scale and/or high-redshift surveys, the physical properties of the sources can only be estimated by exploiting multiwavelength photometry and spectral energy distribution (SED) fitting techniques. These methods involve assumptions that can introduce biases in the estimated physical properties. These assumptions include the choice of an initial mass function (IMF), a stellar population synthesis (SPS) model, and (usually) simple parametric star formation histories. Dust attenuation is modeled as a function of the color excess parameter $E(B-V)$, scaling with the dust column density, and with the interstellar medium (ISM) opacity $k(\lambda)$ which is related to the properties of dust grains (e.g., \citealt{calz94,calz00}). Moreover, fluxes provided by synthetic SED models ought to be corrected for absorption processes in the interstellar medium and intergalactic medium (IGM), which are both wavelength dependent (and the latter is also redshift dependent). These assumptions make the fitting model prone to biases due to the simplifications with respect to the complexity encoded in a real SED (see e.g., \citealt{marchesini09,mob15}).

    To address these challenges, we present FORECAST, a tool for forward modeling cosmological hydrodynamical simulations into mock observed images between rest-frame ultraviolet and near-infrared bands. Unlike existing tools that are primarily galaxy-based (e.g., \citealt{behroozi20,drakos22,snyder22}), FORECAST adopts a particle-based approach, translating the physical properties of individual resolution elements (particles or cells) into observed fluxes. This approach enables the creation of simulated images with realistic galaxy morphologies, interactions, and star formation histories, improving upon standard image simulation software tools, which typically adopt analytical functional forms to render galactic light profiles; readers can refer to GALSIM \citep{rowe15},  \textsc{SkyMaker} \citep{bertin09}, and \textsc{skylens} \citep{plazas19}, for example.

    The use of forward modeling techniques based on numerical simulations is a well-established practice in the literature. They have been employed to assess the reliability of photometric methods \citep{price17,parsotan21} and to evaluate the performance of SED fitting \citep{laigle19}, including the utilization of fully Bayesian inference fitting codes for reconstructing nonparametric star formation histories \citep{ji22}. While many studies often rely on the implementation of phenomenological prescriptions or semi-analytical models to construct mock catalogs \citep{blaizot05,kitzwh07,merson13,bravo20,behroozi20,somerville21,drakos22,yung23}, there is a growing interest on utilizing hydrodynamical simulations to create mock observations with specific scientific purposes \citep{snyder22,cochrane23,barrientos23}.
    When the sample is derived from hydrodynamical simulations, it often comprises a small number of galaxies \citep{guidi16,price17,parsotan21}, or larger samples restricted to a specific redshift range, aimed at simulating observations from specific instruments \citep{sny17,laigle19,snyder22,nanni23}.
    We point out that while some effort has already gone also into building tools that produce images from empirical or semi-analytical models \citep{overzier13,tagh15,bernyk16}, and some final products also publicly available \citep{behroozi20}\footnote{\url{https://www.peterbehroozi.com/data.html}}, the strength of FORECAST stems from its inherently flexible and adaptable framework, specifically designed to emulate real observations and replicate comprehensive photometric surveys by leveraging the predictions of any hydrodynamical cosmological simulation as input.

    The mock images created with FORECAST can be processed and analyzed as real images. As a first application, in this paper we test FORECAST by forward-modeling the \textsc{IllustrisTNG}100 simulation \citep{wein18,pill17,nelson19} and creating a dataset that mimics the observational properties of the GOODS-South field, as observed by the CANDELS survey \citep{grogin11,koek11}, and using CANDELS \citep{guo13} and ASTRODEEP-GS43 catalog \citep{merlin21} for our comparisons. This simulated dataset is publicly available \footnote{\url{http://www.astrodeep.eu/FORECAST}.}, together with a \textit{JWST} CEERS-like dataset and the FORECAST code.

    The paper is organized as follows.
    In Sect. \ref{sc:tool} we describe the methods implemented in FORECAST to forward-model the simulated data.
    The synthetic dataset produced to test the code is described in Sect. \ref{sec:testing}.
    The results of the photometric analysis of our synthetic images are discussed in Sect. \ref{sec:analysis}. 
    In Sect. \ref{sec:release}, we present the public release of the FORECAST code along with an additional dataset of synthetic images emulating the \textit{JWST} CEERS survey. 
    Finally, in Sect. \ref{sec:final} we summarize the main points of our work and discuss possible future work.
    
    All the magnitudes are defined in the AB magnitude system \citep{oke74}, with fluxes in units of $\mu$Jy, namely $m_{AB} = -2.5\,log_{10}(f_{\nu})+23.9$. We adopt the flat $\Lambda$CDM Cosmology constrained by \citet{planck16}, with $\Omega_{\Lambda, 0}$ = 0.6911, $\Omega_{m, 0}$ = 0.3089, $\Omega_{b, 0}$ = 0.0486,  $\sigma_8$ = 0.8159, $n_s$ = 0.9667; and we use the Hubble constant in terms of h $ \equiv H_0/100$ km s$^{-1}$ Mpc$^{-1}$ = 0.6774.
    
    \section{Description of the software}\label{sc:tool}
    
    In this section, we provide a description of the algorithms included in FORECAST to build the synthetic images  (Sect. \ref{sec:observatory}) and to add realistic observational features (Sect. \ref{sec:realism}).

    \subsection{The mock observatory}\label{sec:observatory}
    
    FORECAST uses a particle-based approach to reconstruct the observable photometric properties of galaxies within the field of view. The reconstruction is based on the properties of individual resolution elements in the simulation, represented by stellar particles. Each stellar particle corresponds to a single stellar population (SSP) and collectively forms the simulated objects, representing galaxies in the field of view.\\
    In short, FORECAST reads the physical properties of the stellar particles from the output snapshots of a chosen simulation and translates them into observable quantities as follows. The flux of each particle is computed considering its rest-frame SEDs;
    then, the SED is $k$-corrected consistently with the redshift of the particle. The SED is convolved with chosen passband filter to obtain the theoretical observed flux in that band (Sect. \ref{light}). 
    Gas elements (either particles or cells, depending on the simulation) are used as tracers for dust, which attenuates stellar particle fluxes in the blue and visible range (Sect. \ref{dustatt}). The software does not implement the effects of dust emission.\\
    Finally, the comoving three-dimensional coordinates of each SSP are first projected onto the two-dimensional
    field of view of an observed light cone (Sect. \ref{lc}), and then to a pixel grid (Sect. \ref{map}). Instrumental effects such as PSF blurring and observational noise are added in post-processing (Sect. \ref{sec:realism}).\\ 
    \indent FORECAST has two available options for stellar population synthesis models: \cite{bec03}, modeling stellar emission, and \cite{gutkin16}, which additionally incorporates the rest-frame ultraviolet and optical nebular emission from \textsc{Hii} regions around young stellar populations. 
    We point out that we currently do not include Active Galactic Nuclei and individual Milky Way stars in the rendering of the simulated galaxies; this is left for future work.

    \subsubsection{Input parameters}\label{inpar}
    
    FORECAST is adaptable to the choices of the user by selecting a set of input parameters, described in Table \ref{INpar}. In particular, it is possible to choose the hydrodynamical simulation that provides the backbone of the light cone (box with side-length $L_{box}$); the highest redshift to be included, $z_s$, which determines the maximum distance covered by the light cone, $D_s$; the dimensions of the field of view, $L_{fov}$; the resolution of the ideal simulated images, setting the number of pixels per image side-length $N_{pix}$.

    \begin{table*}[ht!]
        \caption{FORECAST input parameters.}
        \label{INpar}
        \centering
        \begin{tabular}{p{0.15\textwidth}p{0.1\textwidth}p{0.65\textwidth}}
            \hline\hline
            Parameter&Units&Description\\
            \hline
            $L_{box}$&cMpc/h&side-length of simulation box\\
            filters&-&list of N photometric filters (whether provided by the software or by the user)\\
            $L_{fov}$&deg&side-length of the field of view\\
            $N_{pix}$&-&number of pixels per side-length of the mock image\\
            $z_s$&-&highest redshift in the light cone\\
            SED resolution&-&chosen resolution for SED in stellar population synthesis code ("\textsf{lr}" for low resolution, "\textsf{hr}" for high resolution)\\
            IMF&-&chosen IMF for SED in stellar population synthesis code ("\textsf{chab}" for \citealt{ch03} or "\textsf{salp}" for \citealt{salpeter55})\\
            \hline
        \end{tabular}
    \end{table*}
    
    \subsubsection{Light-cone construction}\label{lc}

    Simulation output data are organized in snapshots, which are photographs of the simulated cosmological volume at a specific time of its evolution. FORECAST creates a light cone placing an observer at $z$=0 and rearranging the data from the output snapshots of the chosen simulation, projecting the positions of the simulated objects on a two-dimensional field of view. 

    The light-cone construction procedure is inherited from the software \textsf{MapSim} by \cite{giocoli15}.
    The snapshots used to build the light cone must include the following properties for the stellar resolution elements: 
    \begin{itemize}
        \item comoving coordinates within the simulated volume, ($x_*$,$y_*$,$z_*$) in ckpc
        \item stellar mass, $M_*$ in $M_{\odot}$
        \item initial stellar mass, $M_{i,*}$ in $M_{\odot}$
        \item stellar metallicity $Z_*$ as $M_{Z}/M_{TOT}$
        \item age, $t_{SSP}$ in yr
        \item redshift, $z_*$
        \item subhalo membership ID
    \end{itemize}

    The initial stellar mass is the amount of mass owned by a stellar element when it is born, while the stellar mass accounts for mass returned through winds and supernovae to the ISM by evolved stars.\\
    \indent FORECAST constructs deep light cones stacking simulation boxes, using different snapshots to cover the entire chosen redshift interval (partitions of the light cone). 
    When the length of the simulation box along the $z$-axis, which is the order of a few hundred Megaparsecs in typical cosmological hydrodynamical simulations, is smaller than the distance between two subsequent snapshots, FORECAST adds a replica of the previous or following snapshot, tailoring it to fill the gap. The software replicates the snapshot with the closest redshift to the redshift of the midpoint of the gap.  
    This operation ensures a seamless construction of the complete light cone without any missing sections. 
    Each snapshot is adjusted with rotated, inverted and shifted coordinates to pick structures at an evolutionary stage as close as possible to the one they would be if the entire redshift range was continuously sampled by simulation snapshots.    
    Similar procedures are adopted by \citet{ronc06a}, \citet{croft01} who produced maps to study X-ray emission, and similarly by \citet{scaram93}, \citet{dasilva00}, \citet{dasilva01a}, \citet{dasilva01b} to study the Sunyaev-Zel'dovich effect. 
    
    \begin{figure*}[h]
        \centering
        \includegraphics[width=0.85\textwidth]{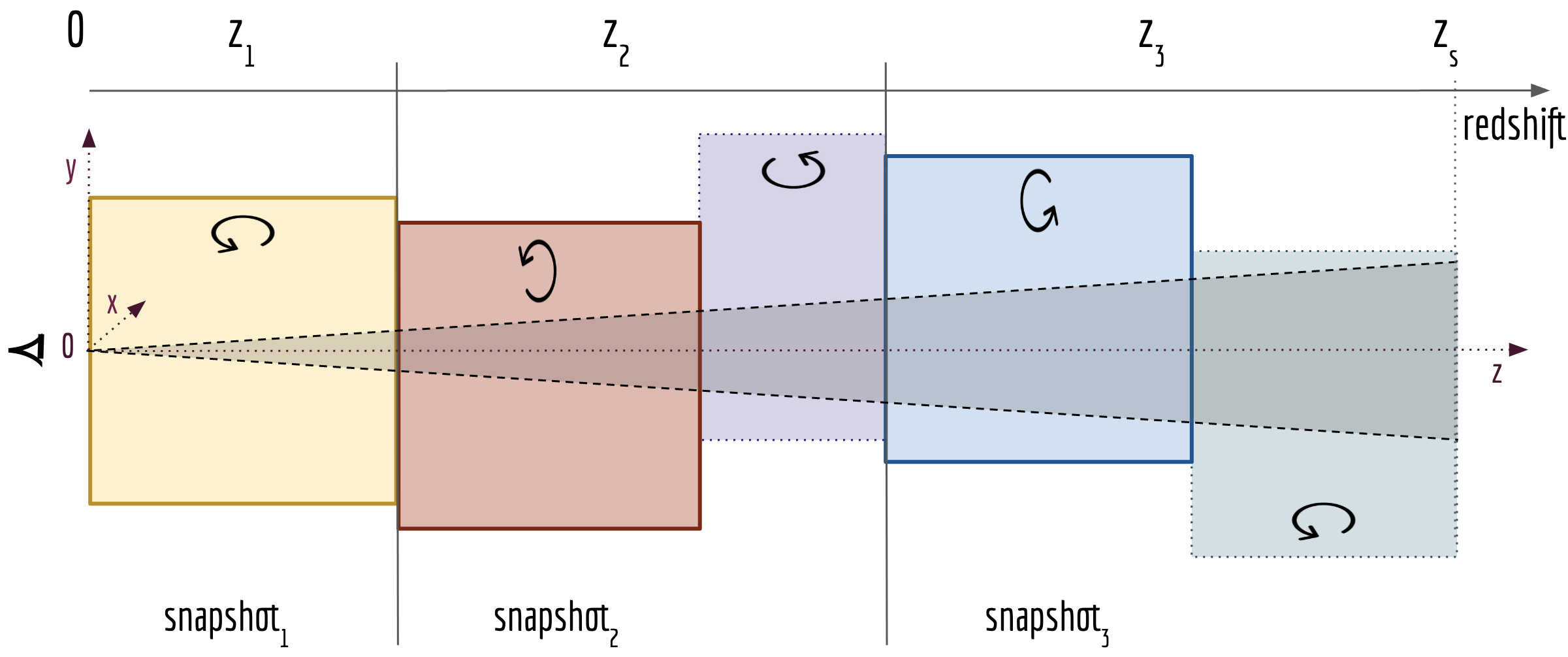}
        \caption{Illustration of the construction of the light cone. The cosmological volume between z=0 and $z_s$ is divided in bins of redshift, delimited by vertical black lines; the observer is located at the position $O$, on the left of the figure, centered with respect to the first simulation box at $z=0$. The full light cone is realized by firstly stacking the comoving volume of the simulation at the proper redshift, represented by snapshots 1, 2, 3 located at redshifts $z_1$, $z_2$, $z_3$ (boxes with thick contours), with rotated coordinates and shifted centers to avoid repetition of the structures along the $z$-axis. Then, the possible gaps between two contiguous snapshots (occurring because the distance between two subsequent snapshots is larger than the size of the simulation box along the $z$-axis) are filled up with replicas of existing snapshots (boxes with dashed contours), with volumes tailored to fill the gaps, and with rotated coordinates and shifted centers. The darker area is the light cone, growing with the comoving distance utmost to the transverse comoving size of the simulation box.}
        \label{planes_cone}
    \end{figure*}
    
    The use of the same snapshot (required to fill potential gaps in the light cone) to reproduce adjacent but different volumes of the cone can cause the repeated appearance of the same structures aligned in radial direction (or in transverse direction, if the same volume is replicated at the same redshift to extend the field of view; this feature is not included). On the other hand, the use of different snapshots to reproduce different volumes of the cone at different cosmic times can also cause the recurrence of objects in the final image because each snapshot of a simulation includes the same sources at different epochs of their evolution. These periodicity effects caused by the repetition of the structures throughout the cone are mitigated by adopting the random combination of the following geometrical readjustments on each of the snapshot boxes used to construct the light cone: (i) the rotation of the positions of stellar particles of $0, \pi/2, \pi$ or $3\pi/2$ around each axis, (ii) the shift of their positions of random amplitude, in [0, $L_{box}$], in (x,y,z) directions, imposing periodic boundary conditions, and (iii) the inversion of one randomly picked axis \citep[see][]{blaizot05}. 
    Figure \ref{planes_cone} shows a sketch of the procedure adopted by FORECAST to construct the light cone, with the colored boxes representing the stacking of the snapshots along $z$, and the darker area outlining the maximum field of view up to $z_s$. 
    
    In a real light cone, the redshift of the sources varies continuously along the line of sight. However, the output of a simulation consists of a finite number of snapshots, each at a given precise redshift - so for example all the particles in the snapshot at $z$=0 have $z$=0, even though the simulation box spans up to many comoving Megaparsecs. To cope with this, the actual redshift assigned to each particle is computed from its comoving distance from the observer (which is computed using its coordinates in the snapshot).\\    
    FORECAST recovers the subhalo membership of each particle, as previously assigned from the simulation procedure (e.g., hydrodynamical simulations usually adopt friends-of-friends group-finding algorithm, \citealt{davis85}; and \textsc{SubFind} algorithm for substructures identification, \citealt{subfind1,subfind2}), in order to track the overall emission of the galaxy.\\
    The software then selects only the particles within the field of view (FoV), whose dimension is assigned by the user in the input file (see $L_{fov}$ in Table \ref{INpar}). To this aim, it computes the distance between the particle and the observer located in the center of the box at $z=0$, that is at (0.5, 0.5, 0.0)$\cdot L_{box}$, and it converts the comoving coordinates of each particle within the cone to angular positions. Particles with right ascension $\alpha_*<L_{fov}$ and declination $\delta_*<L_{fov}$, and with comoving distance $d_*$ within the range of the considered partition ($D_{min} \leq d_* \leq D_{max}$), are included in the cone. The dimension of the FoV cannot exceed the projected angular size given by the comoving box size placed at $D_s$ from the observer. Because the light cone grows to a transverse comoving size equal maximum to the simulation box size, at low redshift only a small region of the simulation box is used. The discontinuities at the edge of the tiled partitions are a standard issue in cone construction (see e.g., \citealt{blaizot05,kitzwh07,bernyk16}). Moreover, some structures might be only partially included and cut on the edge of the field of view; since we work with particles rather than galaxies, we easily identified the partially built structures to be a few percentage ($\sim 0.5 \%$) within the field of view.

    Since the dimension of the input files (snapshots) is typically large and it might be too demanding to have all of them simultaneously saved in the working space, FORECAST is designed to use each snapshot independently, allowing the user to make parallel runs.
    
    \subsubsection{Let there be light}\label{light}
    
    For each stellar particle within the light cone, we then infer observational quantities starting from the knowledge of its intrinsic properties. 
    The \cite{bec03} synthetic stellar population model (\textsc{bc03}) is linked to each stellar particle on the basis of its characteristics, namely the age and the metallicity, assuming a Chabrier \citep{ch03} or Salpeter \citep{salpeter55} initial mass function, at user choice. The stellar particle is assigned to the \textsc{bc03} SED with age and metallicity closest to its nominal age and metallicity.     
    The rest-frame, intrinsic spectral energy distribution $L_{\lambda,*}(\lambda,t,Z)$ of the SSP is then converted into the observer-frame flux per unit wavelength $F_{\lambda,*}(\lambda,t ,Z)$, taking into account the redshift $z$ and therefore the luminosity distance $d_{L,*} (z)$ of the considered particle from the observer:
    
    \begin{equation}
        \label{redflux}
        F_{\lambda,*}(\lambda_{obs},t ,z,Z) = \frac{L_{\lambda,*}(\lambda_{em},t ,Z)}{4 \pi \, (1+z) \, d^2_{L,*} (z)} \, e^{-\tau_{igm}(\lambda_{obs},z)},
    \end{equation}

    \noindent where $\tau_{igm}(\lambda_{obs},z)$ is the optical depth of the intergalactic medium, computed from the IGM absorption model by \cite{inoue14}.
    Finally, the apparent AB magnitude of the SSP at redshift $z$, corresponding to the integrated photon flux collected at $z=0$ from the chosen detector with a filter response $R(\lambda)$ is computed following \cite{fukugita96}. Firstly, FORECAST evaluates the apparent magnitude of a 1 $M_{\odot}$ SSP, namely $m_{\, 1 \, M_{\odot}, \,AB,*}$; it is then rescaled with the initial stellar mass of the particle $M_{i,*}$ in order to follow the same stellar mass loss as in \textsc{bc03}. The final apparent magnitude of the SSP

    \begin{equation}
        m_{AB,*} = m_{\, 1 \, M_{\odot}, \,AB, *} - 2.5 \cdot log_{10} (M_{i,*}/M_{\odot}),
    \end{equation}

    \noindent is finally reconverted into integrated observed flux in units of $\mu$Jy.
    
    \subsubsection{Adding dust attenuation}\label{dustatt}
    
    Correctly taking into account dust extinction by the ISM in a simulation would require the knowledge of the chemical composition, structure, and size distribution of dust grains for each given physical state, and this is rarely included ab initio in simulations. 
    A detailed inclusion of dust physics has only been achieved recently in galaxy formation simulations (e.g., \citealt{bekki15,aoyama18,mckinnon18,graziani20}); more often models incorporate a full treatment of dust with radiative transfer codes (e.g., \textsc{skirt} by \citealt{baes03,baes11}; \textsc{dirty} by \citealt{gordon01,misselt01}; \textsc{sunrise} by \citealt{jonss06, jonss10}; \textsc{hyperion} by \citealt{robit11}), that are able to handle absorption, scattering, and thermal emission by interstellar dust with different solution methods for the radiative transfer equation (e.g., probabilistic methods, numerical methods). However, these methods can be computationally very expensive if the number of particles and/or the number of interactions between particles are increased in the attempt to reduce stochastic fluctuations (e.g., Monte Carlo methods), or can lead to very complex numerical schemes when adopting numerical solutions in the attempt to minimize the introduced numerical errors (e.g., ray-tracing methods).
    
    Since dust resolution elements are not self-consistently included in most large-scale hydrodynamical models and, in general, it is more likely to work on simulations that do not include dust, FORECAST explicitly models the effect of dust in post-processing. It manipulates the properties of gas resolution elements already incorporated in the simulation to turn their observed fluxes, derived considering only the stellar component or the stellar component combined with nebular emission (depending on the model chosen to generate SEDs, see Sect. \ref{sec:observatory}), into dust-corrected fluxes. To improve the readability, we refer to the fluxes computed without the explicit dust attenuation contribution as "dust-free" fluxes, even if they include nebular lines in the modeled SEDs.

    Following \citet{guidr87,devguid00,nelson19,vogel19}, we adopt (i) a semi-analytic model to account for the effect of dust below the resolution limit; (ii) an explicit, geometry-dependent model to account for attenuation by dust in the resolved gas component, using the neutral fraction of gas elements as dust tracer. The fluxes that include these models in their computation are tagged as "dust-corrected" fluxes. We point out that dust emission, including both the predominantly impacting far-infrared and submillimeter wavelengths as well as the mid-infrared contributions from PAH emission \citep{draine21,liu23}, is currently not included in FORECAST.

    For the unresolved dust component, we follow \cite{chfall00}. In their model, young stellar populations ionize the inner regions of dense birth clouds within the ISM; then, line photons emitted in the \textsc{Hii} region, and ultraviolet (UV) and optical nonionizing continuum from young stars are absorbed by dust in the outer \textsc{Hi} region and the ISM. Nonetheless, the stellar UV continuum from stars that are no longer in their birth clouds results to be less attenuated than \textsc{Hii} emission lines  from newborn stars because, after the birth clouds disruption, it is attenuated only within the ISM. Therefore, the intrinsic luminosity of each SSP is obscured as $L_{\lambda,*}^{a}(\lambda)=L_{\lambda,*}^{int}(\lambda) \cdot e^{\tau^{un}_{\lambda}}$, with the unresolved dust optical depth $\tau_{\lambda}^{un}$ 

    \begin{eqnarray}
        \tau_{\lambda}^{un} =
        \begin{cases}
            \; \tau_1 \, (\lambda/5500 \, \text{\r{A}})^{-0.7} & \text{$t_{ssp} \leq$ 10 Myr},\\
            \; \tau_2 \, (\lambda/5500 \, \text{\r{A}} )^{-0.7} & \text{$t_{ssp} >$ 10 Myr}.\\
        \end{cases}
    \end{eqnarray}

    \noindent It is assumed that birth clouds and ambient ISM have the same absorption curves, with different normalization coefficients. All the parameters are taken from the original work, in particular the normalization coefficient at 5500 \r{A}, $\tau_1=1.0$, accounts for both photon lines and continuum radiation absorption; this value is then lowered to $\tau_2=0.3$ after $t_{bc} = 10$ Myr, when birth clouds typically dissipate in Milky Way \citep{murray10,murray11}.\\

    We also include absorption due to resolved dust component by using the distribution of gas-resolution elements in and around each galaxy as tracers for dust (we do not consider intergalactic dust). FORECAST selects gas elements along the line of sight of each stellar particle belonging to a galaxy; by considering the nominal properties of the selected gas along all line of sights, it computes the gas properties averaged over the whole galaxies to determine the dust optical depth. This optical depth is then applied to the full galaxy SED to compute the attenuated galaxy fluxes in the chosen filters. Finally, the software derives the dust attenuation as the ratio between the full galaxy dust-free and the dust-corrected fluxes in each band. This derived dust attenuation is then applied to the dust-free fluxes of each stellar particle in the galaxy, since the software works on a particle-basis.

    \indent In more detail, the dust-corrected (d-c) luminosity of the galaxy is computed from the dust-free (d-f) luminosity, considering the internal dust model by \cite{calz94}, according to which dust and ionized gas are uniformly mixed

    \begin{equation}
        \label{galatt}
        L_{d-c}(\lambda) = L_{d-f} \frac{1-e^{-\tau_{\lambda}^{r}}}{\tau_{\lambda}^{r}},
    \end{equation}

    \noindent  where $\tau_{\lambda}^r$ is the dust optical depth that accounts for absorption and the effect of scattering. Introducing the albedo of grains $\omega_{\lambda}$ \citep{drlee84}, the scattering anisotropy weight parameter $h_{\lambda}$, the scattering optical depth $\tau^s_{\lambda}$, and the absorption optical depth $\tau^{a}_{\lambda}$ for resolved dust elements, we derive the total resolved dust optical depth

    \begin{equation}
        \tau_{\lambda}^{r}= \tau^{a}_{\lambda} \cdot \tau^{s}_{\lambda} = \tau^{a}_{\lambda} \cdot \left[\,h_{\lambda}\sqrt{1-\omega_{\lambda}}+(1-h_{\lambda})(1-\omega_{\lambda})\,\right].
    \end{equation}

    \noindent Scattering parameters are taken from \cite{calz94} in $\lambda$ belonging to [1000,7000] \r{A} range. The resolved dust absorption optical depth \citep{nelson19} depends on the metallicity $Z_g$ and on the neutral hydrogen column density $N_{HI}$ of the gaseous elements as follows

    \begin{equation}
        \label{tau}
        \tau_{\lambda}^{a} = \left(\frac{A_{\lambda}}{A_V}\right)_{\odot} (1+z)^{-0.5} \left( \frac{Z_g}{Z_{\odot}}\right)^{\gamma} \left(\frac{N_{HI}}{N_{HI,0}}\right).
    \end{equation}

    \noindent The first term is the extinction law in the solar neighborhood taken from Table 2 in \cite{cardelli89}. The second and third terms express the dependency of the dust-to-gas ratio on redshift and metallicity, as studied by \cite{dunne11,remi-ruyer14,mckinnon16}. 
    According to \cite{guidr87}, $\gamma$ exponent is a broken power law: $\gamma$=1.35 for $\lambda<$2000 \r{A}, and $\gamma$=1.6 for $\lambda>$2000 \r{A}; the normalization parameters are the solar metallicity $Z_{\odot}=0.02$ \citep{anders89zsolar} and the neutral hydrogen column density $N_{HI,0}= 2.1 \times 10^{21} cm^{-2}$. $N_{HI}$ and $Z_g$ are estimated as \textsc{Hi} mass-weighted averages of the properties of the gas elements that lie along the $z$-axis line of sight of each stellar particle forming a galaxy; $\tau_{\lambda}^{r}$ and $\tau_{\lambda}^{a}$ are computed for the whole galaxy.
    The gas properties are taken from the output files of the hydrodynamical simulation.\newline

    To account for dust in the light cone, FORECAST requires the following properties to be available for each gas particle or cell in the snapshot files:
    \begin{itemize}
        \item comoving coordinates of the geometrical center within the snapshot, ($x_g$,$y_g$,$z_g$) in ckpc
        \item gas mass, $M_g$ in $M_{\odot}$
        \item gas comoving volume, $V_g$ in ckpc$^3$
        \item gas metallicity, $Z_g$ in $M_{Z}/M_{TOT}$ (with $M_Z$ the total mass all metal elements)
        \item neutral hydrogen column density within each gas cell, $N_{HI}$ in $ckpc^{-2}$
        \item redshift, $z_g$
        \item subhalo membership ID
    \end{itemize}
    
    Gas cells are the resolution element in Eulerian hydrodynamical codes or moving-mesh codes (e.g., \citealt{teyssier02,enzo,gizmo,sp10}), and their volume can be approximated as a cube or a sphere. In case of Smoothed Particles Hydrodynamics codes (e.g., \citealt{sp05aG,wadsley04}), where resolution elements are represented by particles, one can consider the smoothing length of gas particles as their spatial extension. FORECAST approximates the volume of a gas resolution element as a cube, thus it computes its linear dimension as $L_g=(V_g)^{1/3}$. \\
    Since we assume Eq. \ref{tau} to relate the properties of the gas with dust absorption, we require $N_{HI}$, whose availability in output may depend on the specific hydrodynamical simulation used to construct the mock images. As an example, for \textsc{IllustrisTNG} the informations about $N_{HI}$ are available only at some redshift; thus we calibrated the relation between the temperature of the gas cells and the ratio between the neutral hydrogen mass and the total mass within gas cells on TNG data to evaluate the neutral hydrogen column density, as we show in Sect. \ref{sec:TNG}. 
    
    For each stellar particle, FORECAST selects the gas resolution elements that have its same subhalo membership, satisfying the following conditions along its $z$-axis line of sight

    \begin{equation}
        \begin{cases}
            \, x_g - 0.5 \, \cdot L_g \leq x_* \leq x_g + 0.5 \, \cdot L_g,\\
            \, y_g - 0.5 \, \cdot L_g \leq y_* \leq y_g + 0.5 \, \cdot L_g,\\
            \, z_g < z*.
        \end{cases}
    \end{equation}

\begin{figure}[h!]
        \centering
        \includegraphics[width=0.45\textwidth]{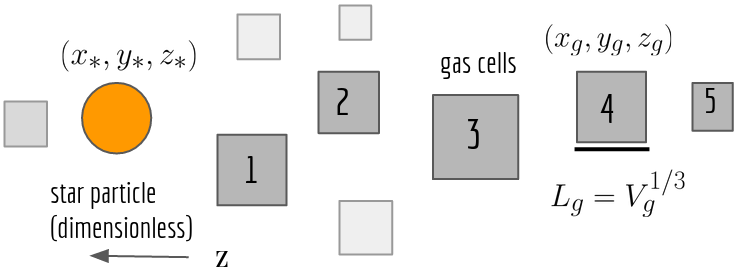}
        \caption{Selection procedure of gas cells (gray squares) along the line of sight of a stellar particle (yellow circle). The stellar particle, with comoving coordinates ($x_*$, $y_*$, $z_*$), is a dimensionless point, while gas cells have a linear extension $L_g$ in (x,y,z), assuming cubic cells with volume $V_g$. The $z$-axis grows from right to left, meaning that darker gray gas cells are in front of the star particle and along its line of sight (LoS), compared to the observer located at the origin n of the $z$-axis. Lighter gray gas cells are not included in the selection, because they are behind or out of the LoS of the star particle. }
        \label{los}
    \end{figure}

    The selection procedure is illustrated in Fig. \ref{los}, in case gas resolution elements are cells.

    To maintain the computing time reasonable (each snapshot might include up to billions of star and gas elements), we compute the dust optical depth on the whole galactic spectrum by averaging the properties of the gas selected over the entire galaxy. Then, we derive and assign the attenuation in each band to the fluxes of each SSP constituting the galaxy.\\    
    The neutral hydrogen column density and the metallicity of the gas within each galaxy are independently computed as the result of neutral hydrogen mass-weighted quantities of the selected i-th gas cells in front of each SSP belonging to that galaxy as follows 

    \begin{equation}
        \begin{split}
        <Z_{g}>_{gal} =& \frac{\sum_i  Z_{i,g} \cdot M_{i,HI}}{\sum_i M_{i,HI}},\\
        <N_{HI}>_{gal} =& \frac{\sum_i  N_{i,HI} \cdot M_{i,HI}}{\sum_i M_{i,HI}}.
        \end{split}
    \end{equation}

    \noindent These quantities, which are computed on galaxy basis, are used to estimate the resolved dust optical depth in Eq. \ref{tau}, which is applied to each galaxy SED; then, we determine the galaxy dust-corrected observed flux per unit of wavelength, which is then evaluated within the chosen filter response $R(\lambda)$ to obtain the galaxy dust-corrected integrated flux.

    Galaxy dust-free integrated flux and dust-corrected integrated flux are finally used to estimate the average resolved dust attenuation for each galaxy

    \begin{equation}
        \begin{split}
            A^{r}_{gal} =& \, -2.5 \, log_{10} \, \left( \frac{F_{gal,d-c}}{F_{gal,d-f}} \right) = \, -2.5 \,log_{10} \, \left(\frac{1-e^{-\tau^{r}_{\lambda}}}{\tau^{r}_{\lambda}} \right).
        \end{split}
    \end{equation}

    \noindent Since we use a particle-based approach and we are interested in obtaining the observed integrated flux of each SSP in the light cone, this attenuation is then applied to the dust-free apparent AB magnitude of each stellar particle, to get their dust-corrected apparent AB magnitudes

    \begin{equation}
    m_{AB,*}^{d-c} = m_{AB,*}^{d-f} + A^{r}_{gal}.
    \end{equation}

    \noindent To map the light of the SSPs onto the mock image, we convert the dust-corrected apparent AB magnitude of each SSP into observed flux, in $\mu$Jy units. We point out that to streamline computational time, our approach assumes uniform dust attenuation for the SSPs belonging to the same galaxy. Nevertheless, we are working on enhancing both the computational efficiency and the realism of the dust attenuation model to achieve more accurate results.
    
    We accurately checked that the adopted dust model produces the expected changes in the observed fluxes: accounting for dust, the rest-fame UV and optical magnitudes are the most affected, increasing up to one order of magnitude with respect to their dust-free counterpart, especially in shorter wavelengths, while IR magnitudes remain unchanged.

    \subsubsection{Mapping}\label{map}
    
    The fluxes of the stellar particles that build up the simulated galaxies must finally be mapped on a bi-dimensional pixel grid. First, the angular coordinates of the particles in each partition (portion of the light cone) are projected on a bi-dimensional plane located in the central point of the considered partition, and then they are translated in pixel coordinates.\\ 
    FORECAST first converts the right ascension and declination coordinates of the particles in the considered partition into pixel coordinates, accounting for the FoV dimension in pixels ($N_{pix}$ is defined in Table \ref{INpar}).
    The total flux of each pixel is obtained as the sum of the fluxes of all particles having coordinates within it. 
    Since, at this stage, the synthetic image is ideal (i.e., the number of photons hitting the mirror of the telescope is a smooth function from the theoretical intensity of any source; there are no diffraction effects due to the limited surface of the optics of the telescope, or other sources of uncertainties and errors), we choose not to apply any kernel convolution matrix that distributes the flux of a particle on adjacent pixels; instrumental effects, including noise and PSF smoothing, are attached in post-processing on the final image in a chosen filter (see Sect. \ref{sec:realism}). In the real world, each single point source (i.e., SSP) gets smeared by the PSF of the telescope, and therefore the most accurate way to simulate this effect would be to stack PSF stamps (one per SSP), shifted, and rebinned in order to have their centers at the exact, sub-pixel position of the corresponding particle. However, we have implemented this algorithm and checked that the final result is virtually identical to another one in which we first sum the flux of all the SSPs falling into a pixel, and then simply PSF-smooth the light of that pixel. We therefore choose to adopt the latter method, which is simpler and requires a much smaller expense of computational time. 
    
    The final image is obtained by stacking the images corresponding to each projected plane of the partitions that build up the light cone.

    \subsection{Adding realism: Noise and PSF}\label{sec:realism}
    
    Here we describe the two final steps necessary to simulate a real mosaic, which are the inclusion of observational noise and the convolution with a Point Spread Function of the sources of interest. These two steps are independently performed with a \textsf{python} script and can be executed multiple times on the ideal images produced by FORECAST to simulate them with different depths. 
    
    When FORECAST assembles its output in a chosen band, the simulated image is noiseless and its resolution is only limited by the pixel scale. 
    First of all, as mentioned in Sect. \ref{sec:construction}, we convolve it with the PSF of the instrument and filter that is being simulated; this spreads the flux coming from a single pixel over an extended region.  
    
    Then, we add a noise background that limits the depth of the image. To this aim, we create an RMS map as a flat image with a constant value, that is the chosen standard deviation of the background noise pixels, which can be obtained from the desired limiting magnitude of the image as
    
    \begin{equation}
        \sigma = \frac{10^{-0.4(m_{lim,n}-ZP)}}{n \sqrt{\pi}r_{ap}} \label{sigma},
    \end{equation}
    
    \noindent where $n$ is the signal-to-noise ratio to which the limiting magnitude corresponds. $r_{ap}$ is the radius of the aperture used to compute the limiting magnitude; it does not have a fixed value, as it depends on how the depth of an image is defined (e.g., if a band has limiting magnitude $mag_{lim}$ at N$\sigma$ in 2", the radius will be 1" in pixels, and the noise level will be set to match the expected depth). ZP is the zero-point of the image (in this case, $ZP=23.9$ with the pixels in $\mu$Jy).

    Then, we add to the RMS map the contribution of the photon noise. This is done by replacing the value of each pixel, $\sigma_{sky}$, with 

    \begin{equation}
        \sigma_{tot}=\sqrt{\sigma^2_{sky}+\frac{f_{source}}{t_{exp}}},
    \end{equation}

    \noindent where $t_{exp}$ is the total exposure time of the image, and $f_{source}$ is the flux coming from luminous sources falling in that pixel (this formula can be derived from first principles and is discussed in \citealt{merlin22euclid}).
    
    Finally, we create the noise image as a random  realization with the $\sigma$ of the distribution for each pixel provided by the RMS map, and we sum it to the original noiseless image containing the simulated galaxies to obtain the final mock image. The RMS map can also be used to compute the errors on any photometric measurement performed on the scientific image.
    
    We included an option to further slightly smooth the scientific image, in order to simulate the result of possible noise correlation of pixels introduced by mosaicing, rebinning, and stacking. The apparent limiting magnitude of the scientific image (i.e., the noise standard deviation when the background is subtracted) can be chosen to be different from the one given by the RMS map; if this option is chosen, the image is iteratively smoothed with a gaussian kernel until the desired apparent depth is obtained.

    All the images are finally normalized to $\mu$Jy units by default, meaning that the magnitude associated to each pixel flux value is simply $m=-2.5\mbox{log}(f)+23.9$. We include the possibility of creating the images with any different normalization, giving a different zero-point in input. The values that can be configured to perform the post-processing of the image are summarized in Table \ref{t:postproc_params}. 

    \begin{table}[ht!]   
        \caption{Parameters for image post-processing.}
        \label{t:postproc_params}
        \centering
        
        \begin{tabular}[t]{ccc}
            \hline\hline
            Parameter&Units&Description\\
            \hline
            \rule{0pt}{2ex} 
            INPUT imm& - &FORECAST image\\
            PSF& - &Point Spread Function\\
            ZP& - &image zero-point\\
            FWHM& arcsec &full width half maximum\\
            TEXP& s &integrated exposure time\\
            RMSimg& - &\\
            RMSerr& - &\\
            PS& arcsec/pixel & pixel scale\\
            APER & arcsec & diameter aperture\\
            BKGD& - & background\\
            \hline
        \end{tabular}
    \end{table}
    \section{Testing FORECAST: Emulation of the CANDELS GOODS-South Field}\label{sec:testing}

    Before exploiting the tool to make forecasts for the next-generation surveys, we tested its capabilities emulating a well-known dataset, to compare the new mock data with existing photometric catalogs, investigating which are the most relevant tensions with observed data, and how the procedure can be improved for future work. These data products, together with the \textit{JWST} CEERS dataset (see Sect. \ref{jwst}), are publicly available for scientific analysis.
 
    In Sect. \ref{sec:construction}, we illustrate the procedure adopted to build the images and catalog analyzed in this work. We describe the cosmological hydrodynamical simulation employed to build the images in Sect. \ref{sec:TNG}. We then describe the mock light-cone setup adopted for the present analysis and give an overview of the simulated images in Sect. \ref{setup}. 
      
    \subsection{Constructing the images}\label{sec:construction}
    
    In order to examine the performance of the software tool, we tested the synthetic images against a thoroughly investigated observational counterpart. To this aim, the size and filter set of the simulated FoV and the extension in redshift of the mock light cone have been chosen to emulate the Great Observatories Origins Deep Survey Field South (GOODS-South, GS), making use of the state-of-the-art hydrodynamical simulation \textsc{IllustrisTNG}. The GS field, located at RA=3h 32m 30.39s and Dec=-27$^{\circ}$ 48m 11.28s with a covered region of 10' $\times$ 16', has been targeted for deep, multiwavelength observations from ground and from space as part of several survey programs (e.g., \citealt{grogin11, brammer12,curtis-lake22,robertson22}).
    
    We elect to produce thirteen images corresponding to thirteen band-passes covering from the rest-frame optical to the near-infrared (NIR) wavelength range: $HST$ ACS $B$435, $V$606, $I$814, $Z$850, and WFC3 $Y$105, $J$125, $JH$140, and $H$160; plus a ground-based $VLT$ HAWK-I $Ks$ band, and four IRAC channels, namely CH1, CH2, CH3, and CH4. We post-processed the FORECAST ideal images with the real PSF models adopted in the CANDELS image analysis described in \cite{guo13}, see Sect. \ref{sec:realism} for further details. 
    Since our goal was to test the tool over a fairly wide range of wavelengths, we only emulated 13 out of the 17 bands of the reference catalog.

    \subsubsection{IllustrisTNG simulation}\label{sec:TNG}
    
    The \textsc{IllustrisTNG} (or TNG) Project \citep{wein18,pill17,marinacci18,naiman18,nelson18,sp17} is a suite of cosmological magneto-hydrodynamical simulations, following the evolution of cosmological volumes between $z = 20$ to $z = 0$. The moving-mesh code \textsc{arepo} \citep{sp10} solves coupled equations for gravity and magneto-hydrodynamics: Poisson's equations for full Newtonian gravity are treated with a hybrid TreePM scheme \citep{treepm2,treePM}, while an unstructured and moving Voronoi mesh is adopted to solve equations of hydrodynamics. Processes such as the cosmic gas accretion into halos, tidal and ram-pressure stripping, and dynamical friction, as well as the hierarchical growth of halos and galaxies, and galaxy mergers naturally emerge as the solution of the equations of gravity and hydrodynamics in an expanding Universe with gravitationally collapsing structures.\\
    Its galaxy formation model (see \citealt{wein18,pill17}) is built upon the original \textsc{Illustris} model \citep{vogel13,genel14,sij15} and accounts for all the processes occurring below the resolution scale of the simulation. It includes gas density-threshold
    star formation, adopting a Chabrier IMF \citep{ch03}, and evolution of stellar populations represented by star particles; chemical enrichment of the ISM with the tracking of nine chemical elements (H, He, C, N, O, Ne, Mg, Si, Fe); gas heating and cooling; feedback from supernovae through galactic winds; seeding and growth of supermassive black holes, and energy- and momentum-driven feedback into the surrounding gas.
    TNG cosmology is consistent with recent observational constraints from \citet{planck16} (cosmological constant $\Omega_{\Lambda, 0}$ = 0.6911, matter density $\Omega_{m, 0}$ = 0.3089, baryon density $\Omega_{b, 0}$ = 0.0486, power spectrum normalization  $\sigma_8$ = 0.8159, and spectral index $n_s$ = 0.9667; the Hubble constant is $H=100 h$ Mpc, with $h$ = 0.6774).
    
    The \textsc{IllustrisTNG} Project consists of three physical simulation boxes with periodic cubic volumes of roughly 50, 100, and 300 comoving Mpc (cMpc) side-length, named \textsc{TNG50}, \textsc{TNG100}, and \textsc{TNG300} respectively, each of them reproduced at high, medium and low-resolution level (-1, -2, -3 suffix). Gas cell masses (i.e., the mass resolution of a simulation) and gas cell sizes (i.e., the spatial resolution of a simulation) both form continuous distributions. Stars inherit the gas mass from which they form, so they also have a variable mass resolution, and they continuously decrease in mass due to stellar evolution \citep{pill17}.\\ 
    The \textsc{IllustrisTNG} dataset has been publicly released with a complete user guide in \cite{nelson19}.
    In each output snapshot of the simulation, overdensities of dark matter are identified using the Friends-of-Friends algorithm \citep{davis85}; self-bound subhalos, which are primarily constructed with dark matter particles, and have their baryons (i.e., gas and stars) associated with the geometrically nearest DM particle, are later identified using the \textsc{SubFind} algorithm for substructures identification \citep{subfind1,subfind2}. 
    They also generate two distinct merger trees at the subhalo level, with \textsc{SubLink} \citep{rg15} and \textsc{LHaloTree} \citep{sp05} algorithms.
    
    This work makes use of the TNG100-1 simulation boxes ($L$= 75 $h^{-1}$Mpc = 110.7 Mpc), but the tool is configured to use any realization of the simulation.
    
    \subsubsection{Emulation setup and output}\label{setup}
    
    We created a light cone from $z=0.1$ (we excluded lower redshift snapshot to avoid excessive contamination from large local sources) up to $z_s=7.2$, corresponding to a comoving distance $D_s = 6034.14$ cMpc $h^{-1}$ (with $h$=0.6774). The light cone is built with 122 partitions, with 89 output snapshots. We adopt a Chabrier IMF for the modeled SSPs, consistently with the TNG choice.
    
    The mock survey is designed to provide a coverage of galaxies from optical to near-IR wavelengths, emulating a squared field of view of 200 sq. arcmin (comparable to the GOODS-South field area) realized on a grid of 200 million pixels, resulting in a pixel scale of 0.06 arcsec, which is the typical for \textit{HST} observations (for simplicity, the \textit{VLT} image and the \textit{Spitzer} images were directly created on the same pixel scale, rather than going through a rebinning process).\\  
    We simulated the thirteen broad-band images adopting resolution and limit magnitude from \cite{merlin21}, shown in Table \ref{filterReD}.

    \begin{table}[ht!]
        \caption{Summary of the instrumental PSF and depths adopted for the image simulations in this work from \cite{merlin21}.}
        \label{filterReD}
        \centering
        \begin{threeparttable}
        \begin{tabular}[t]{cccc}
            \hline\hline
            Instrument&Filter&PSF(arcsec)&5$\sigma$ depth AB\\
            \hline
            \textit{HST} ACS&F435W&0.08&$\text{28.83}\tablefootmark{a}$\\
            \,\,\,\,&F606W&0.08&$\text{29.24}\tablefootmark{a}$\\
            &F814W&0.09&$\text{29.35}\tablefootmark{a}$\\
            &F850LP&0.09&$\text{28.54}\tablefootmark{a}$\\
            \textit{HST} WFC3&F105W&0.15&$\text{28.70}\tablefootmark{a}$\\
            &F125W&0.16&$\text{28.85}\tablefootmark{a}$\\
            &F140W&0.17&$\text{27.64}\tablefootmark{a}$\\
            &F160W&0.17&$\text{28.72}\tablefootmark{a}$\\
            \textit{VLT} HAWK-I&$\text{K}_s$&0.43&$\text{26.26}\tablefootmark{b}$\\
            \textit{Spitzer} IRAC&CH1&1.66&$\text{25.63}\tablefootmark{b}$\\
            &CH2&1.72&$\text{25.51}\tablefootmark{b}$\\
            &CH3&1.88&$\text{23.28}\tablefootmark{b}$\\
            &CH4&1.98&$\text{23.16}\tablefootmark{b}$\\
            \hline
        \end{tabular}
        \tablefoot{
            \tablefoottext{a} {Median aperture magnitudes within a fixed radius of 0.17”; the given values are averages of the varying depths in the CANDELS GOODS-South field (including CANDELS-deep field and the HUDF depths).\\}
            \tablefoottext{b} {Median total magnitude at 5$\sigma$; the given values are averages of the varying depths in the CANDELS GOODS-South field.}}
        \end{threeparttable}
    \end{table}
    
     \begin{figure*}[h!]
        \centering
        \includegraphics[width=0.75\textwidth]{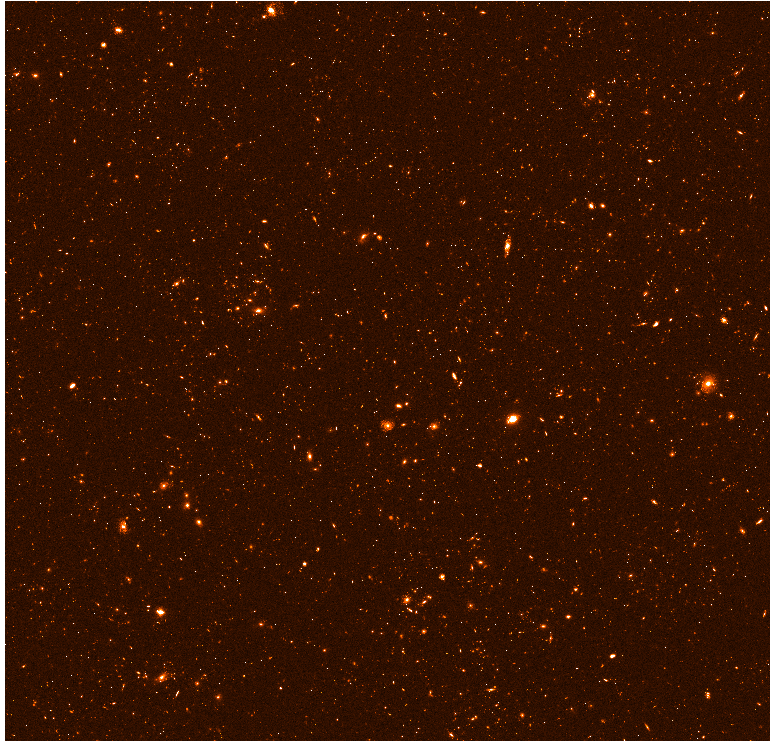}
        \caption{Final simulated image in $H$160 band, after post-processing with PSF and noise, color-coded by fluxes in units of $\mu$J.}
        \label{Hfull}
    \end{figure*}     
    
\begin{figure*}[h]
        \centering
        \includegraphics[width=0.6\textwidth]{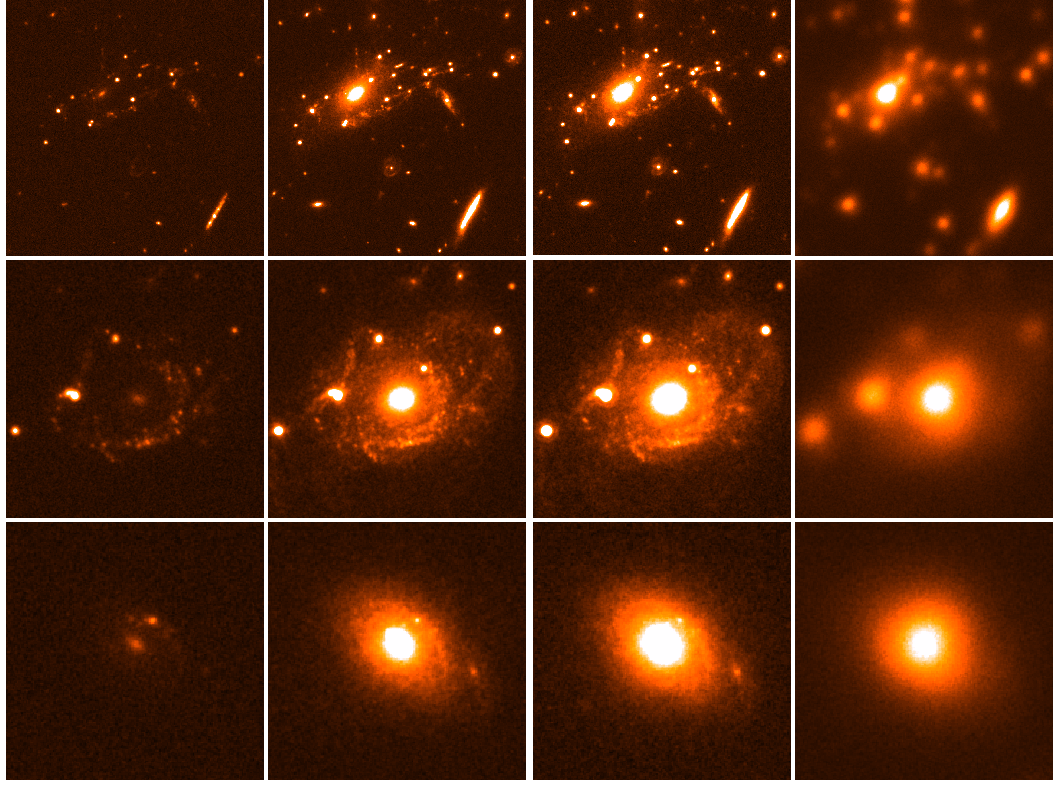}
        \caption{\label{ALLbands1} Three examples of small areas of the simulated field of view containing a small group of galaxies or single objects (top to bottom), in four simulated bands (left to right: $B$435, $Y$105, $H$160, IRAC CH1, in $\mu$J units). They are low-redshift sources, with $z$=0.35-0.45. The sizes of the areas are 0.401, 0.062 and 0.027 sq. arcsec, from top to bottom.}
    \end{figure*}

    We show the final simulated F160W (or $H160$) image, post-processed with observational features, in Fig. \ref{Hfull}.
    The complex morphologies and interactions of galaxies as observed in the real sky beautifully show up in the simulated image.
    We show three examples of galaxies or small groups located in the final FoV in four of the thirteen simulated bands in Fig. \ref{ALLbands1}. These are all located in the low-redshift Universe, between $z$=0.35-0.45. It is possible to appreciate the morphology and brightness changes across the spectrum, with sources appearing more luminous in infrared bands because of the typical SED shape of galaxies in which the star formation activity is not prominent.
    Bluer bands show more clearly the signs of recent star formation activity as concentrated blobs of high luminosity. It is possible to fully characterize these regions by checking the true ages and metallicities of the corresponding SSPs. 

    \subsection{Photometric analysis}\label{sec:analysis}
    In order to validate the accuracy of the simulated field, we proceeded using a processing pipeline that is very similar to the one typically used to extract the photometric information, and then the scientific properties, from real imaging data. We have only bypassed the reduction processing steps of a typical raw imaging dataset (e.g., flat-fielding, bias and background subtraction, mosaicing, etc.), assuming they have been performed in an ideal way. We also did not extract new PSF models from the images, exploiting the ones used to build the simulated images.
    
    Therefore, we start our analysis from a simulated image that is comparable to the final mosaic on which the CANDELS team performed the photometric measurements for the final catalogs. Specifically, for our analysis and comparisons, we used our multiwavelength photometric catalog ASTRODEEP-GS43, which is an upgrade of the CANDELS catalog by \cite{guo13}, providing photometric fluxes in 43 passbands, plus physical properties and estimations of the photometric redshift for $\sim$ 35,000 sources located in the GS field.
    
    The final output of FORECAST used for the analysis are (i) the synthetic images in all the simulated filters (see Table \ref{filterReD} for resolution and depth informations), post-processed as described in Sect. \ref{sec:realism}, and (ii) the input galaxy catalog (see Table \ref{galaxycat} for details on the fields contained in the catalog), that we call Input Universe (IU).
    This catalog, which is galaxy-based, is built from the properties of the stellar particles listed in the \textsc{IllustrisTNG} output and included in the light cone by FORECAST. The additive properties (e.g., mass, flux) are computed as the sum from all SSPs belonging to a given galactic subhalo in the simulation; metallicity and age are weighted with the stellar mass of the SSPs, and redshift is computed as the mean value of the redshifts of all the membership SSPs. The coordinates of the center of the galaxies are computed as the flux-weighted sum (in $H$160) of the coordinates of all the particles of the subhalo. For extended objects, which appear as separate clumps of light in the image but are identified as single objects in the \textsc{IllustrisTNG} simulation, we refine the center computation with a 3$\sigma$-clipping procedure; that is, we only consider the SSPs for which the distance from the previously determined center R($x_i$,$y_i$,$x_{c,0}$,$y_{c,0}$)<3$\sigma(R)$, thus excluding scattered, isolated particles which might bias the estimate of the center coordinates.
    
   \subsubsection{Detection and photometry}\label{detandphot}
   
    We performed the photometric analysis on the simulated images following the procedure adopted by \cite{guo13} and \cite{merlin21}. We first detected sources on the $H$ band image with \textsc{SExtractor} \citep{bert91} adopting Hot+Cold detection for a finer deblending of the sources. We then remeasured $H$ fluxes with \textsc{a-phot} \citep{merlin19aphot}, which yields a less biased estimate of the total flux (see \citealt{merlin22}). Fluxes in the remaining $HST$ bands were obtained by correcting the total $H$ flux by the color aperture term of the considered band, that is $f_{band,tot}=f_{H,tot}\times (f_{band,segm}/f_{H,segm})$, with the fluxes again measured with \textsc{a-phot}, after PSF-matching all the images to the $H160$ resolution as described in \cite{guo13}.\,
    For the $K$ band and the four IRAC bands, which have lower resolution than the detection $H$ image, we used \textsc{t-phot} \citep{merlin15,merlin16b} to perform template-fitting photometry; we also measured $H$ band fluxes on the image PSF-matched with the lower-resolution ones, again to estimate a robust color term.
    Finally, we assembled a catalog with the total fluxes of all detected sources in all bands. All the uncertainties associated with the flux measurements have been computed using the RMS maps created as described in Sect. \ref{sec:realism}.

    \subsubsection{Multiband photometry}\label{sec:mbp}

\begin{figure*}[h]
        \centering
        \resizebox{\hsize}{!}{\includegraphics{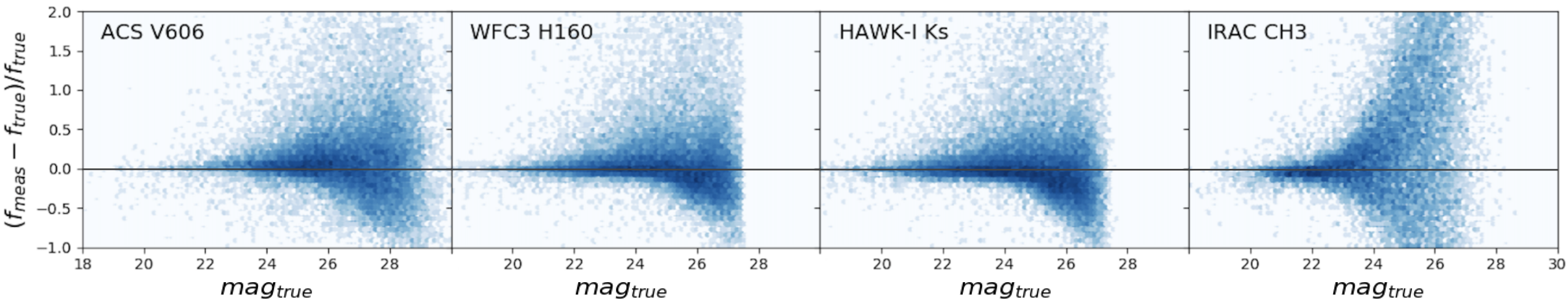}}
        \caption{\label{phot} Comparison between input and measured fluxes of the detected galaxies matched with the IU, in the $V$606, $H$160, $K_s$, and IRAC CH3 simulated band. The panels show the relative errors in flux measurement, $(f_{meas}-f_{true})/f_{true}$, as a function of the input true magnitude; so values above zero mean over-estimation of fluxes with respect to the true values, while values below zero mean under-estimation. The color coding gives the density of sources in every point of the plots.}
    \end{figure*}

    We assessed the accuracy of the fluxes measured with the standard photometric approach by comparing them with the input, true fluxes. The latter can be easily obtained as the sum of the dust and IGM-attenuated flux from all SSPs belonging to a given galactic subhalo in the simulation. 
    Figure \ref{phot} shows the comparison between the input fluxes and the fluxes measured in $V$606, $H160$, $K_s$ and IRAC CH3 bands, after a spatial cross-correlation of the \textsc{SExtractor} detections with the sources in the input IU catalog (only considering galaxies with $H_{true}<27.5$) using \textsc{TopCat} \citep{taylor05} to find the closest neighbors within a searching radius of 3 FWHM. We found 32,413 matched objects (99,6$\%$ of the total; the missing ones are spurious detections). We also checked the other nine simulated bands. The overall agreement is good. 
    The most prominent features are the presence of some bright outliers, most likely due to contamination from neighboring sources, and most of all, a declining trend at low magnitudes; we note that the trend is present in the detection band $H$160, on which the total flux used to scale the colors in all bands is computed, and it is propagated to all other bands, while the colors term are estimated robustly.

    \subsubsection{H160 number counts}

\begin{figure}[h]
        \centering
        \includegraphics[width=0.50\textwidth]{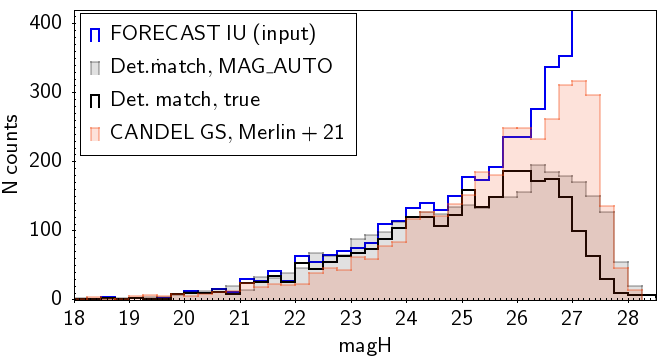}
        \caption{Number counts in the $H$ detection band, relatively to an area of $\sim$16 sq. arcmin in order to be comparable with a region of the same area and with homogeneous depth from CANDELS GOODS-South DEEP (red shaded histogram). Blue line: IU counts; gray shadow: all detections, using \textsc{SExtractor} \texttt{MAG\_AUTO} as total flux estimate; black solid line: IU (true) flux of the detected sources with a IU match.}
        \label{Hcountsnew} 
    \end{figure}

    We then checked the number counts as a function of the magnitude of the detected objects in the simulated $H$ image. 
    Figure \ref{Hcountsnew} shows a comparison of the counts  between the simulation (both the IU and the detections) and a sample of the CANDELS GS "Deep" area. For this comparison, the GS sources were selected taking a crop of 4000$\times$4000 pixels ($\sim$16 sq. arcmin) in the deep region (centered on RA 53.0899 and Dec -27.8050), thus avoiding the Hubble Deep Field and the shallower "Wide" region. We considered a region of the simulated $H$ image with an equal area, and not containing large local objects, to compare the results fairly. After a cross-correlation between the coordinates of the sources in the IU and those in the catalogs of the detection, we found 3,130 sources in the considered area out of 35,459 detected sources in the full FoV.
    
    The comparison shows good agreement between the IU (blue line) and CANDELS (red shaded area) down to magnitude $H\sim26$, after which the CANDELS counts start to deviate from IU, peaking at $H\sim27$ and falling at fainter magnitudes because of the incompleteness of the observed catalog.   
    We also show two cases for the counts of the sources detected on the simulated image: the \textsc{SExtractor} measured \texttt{MAG\_AUTO} of all detected sources that have a match in the IU catalog (gray shaded area); and the true $H$ magnitude of the same sources (black solid line). The distribution of the detected sources with true magnitudes is brighter with respect to the same distribution with measured values, as expected, since some flux in the faintest wings of the light profiles is always lost if a finite aperture is used to measure it. With the present configuration, the counts of the detected sources in the simulated image depart from IU counts at $H\sim25$ and peak at $H\sim26.5$, slightly brighter than the peak at 27.0 of CANDELS counts. What is more interesting, though, is that the detections in the simulated image are well below the CANDELS counts already at $H\sim26$.
    
    We tried to pinpoint the origin of this discrepancy in the number counts at 26<$H$<27. To check whether these inconsistencies could be due to an unfortunately sub-dense realization of the simulation (since the reference light cone is constructed sampling a random area of each snapshot, see Sect. \ref{lc}), we built and analyzed a tailored light cone selecting regions of the \textsc{TNG} snapshots having stellar mass density close to the average of the full simulated volume. However, the resulting counts were substantially similar, both in the IU and in the detections.  
    
    Then, to check whether the problem could originate from the input catalog or from the detection procedure, we created a new realization of our $H$ image, including only sources with true magnitude in the relevant range, to avoid contamination and blending with brighter sources.
    We compared this tailored simulated image (which we tag as "reference IU") with a similar image simulated with GALSIM \citep{rowe15}, which we fed with a mock galaxy catalog produced with the EGG software \citep{schreiber17}, again only including galaxies of the same input magnitude range. EGG catalogs are based on empirical relations calibrated on the observed CANDELS data, extrapolated to faint magnitudes; we created a catalog with a faint limiting magnitude ($H=31$) to ensure input completeness in our considered magnitude range.
    After having applied our post-processing pipeline, we run \textsc{SExtractor} with the very same parameters adopted for the detection on the reference image. 

    \begin{table}   
        \caption{Number of galaxies with 26<$H$<27 in reference (Ref.) IU and EGG/GALSIM IU; their respective number of detected sources (Det.); and the detected sources matched with their respective IUs (match).}
        \label{t:Ngals}
        \centering

        \begin{tabular}[t]{ccc}
            \hline\hline
            sample 26<$H$<27&$N_{gal}$ Ref.&$N_{gal}$ GALSIM\\
            \hline
            \rule{0pt}{2ex} 
            IU&12,829&14,883\\
            Detected &11,318&14,514\\
            Matched &11,135&14,299\\
            \hline
        \end{tabular}
    \end{table}

    As shown in Table \ref{t:Ngals}, the reference IU is less populated than the EGG IU, counting $\sim$ 13,8$\%$ fewer galaxies.
    On top of this, $\sim 1500$ true sources are not detected in the FORECAST image, against only $\sim 370$ missed on the GALSIM image. 
    
    We investigated the nature of unmatched undetected sources in both samples, distinguishing between sources that are not detected because they are blended with, or obscured by, other objects, and sources that may have very low surface brightness. 
    Looking at the undetected sources in the reference IU (1694 sources), we found that 38$\%$ (638) is composed of blended sources, which are objects falling within larger and brighter galaxies in the full image; the remaining 62$\%$ (1061) consists of isolated galaxies, which can be either sources with low surface brightness or objects fragmented in multiple conglomerates of light which are individually too faint to be detected, the latter being a common kind of object generated by hydrodynamical models. In the EGG IU, the unmatched undetected sources are $51\%$ (298) blended with close galaxies, and the remaining $49\%$ (286) composed of isolated galaxies.
    
    We conclude that the deficiency of detections in the FORECAST simulated image at $26<H<27$ is due to two main factors: firstly, the IU is ab initio less dense than one generated using empirical prescriptions, possibly implying that the \textsc{IllustrisTNG} universe contains less faint galaxies than expected; then, a fraction of objects is undetected because of blendings and superpositions (which is reasonable given that it is very difficult to identify faint objects obscured by brighter ones in the real sky), and because of their fragmented morphologies, which do not show up in GALSIM galaxies given the analytic light profiles from which they are generated.
    We will investigate further this issue in future work.

    \subsubsection{BzK diagram} 
    The $BzK$ diagram \citep{daddi04} is a widely used diagnostic color-color plot, useful to separate star-forming and quiescent galaxies using the observed $B - z$ and $z - K$ colors of $1.4 \leq z \leq 2.5$ sources. The criterion is empirical, based on the spectroscopic redshifts from the K20 survey \citep{cimatti02} and other publicly available data sets; however, synthetic stellar populations of both kinds (i.e., star-forming and passive) have been shown to indeed occupy the corresponding areas in this plot, when redshifted to $1.4 < z < 2.5$. 
    
\begin{figure}[h]
        \centering
        \includegraphics[width=0.50\textwidth]{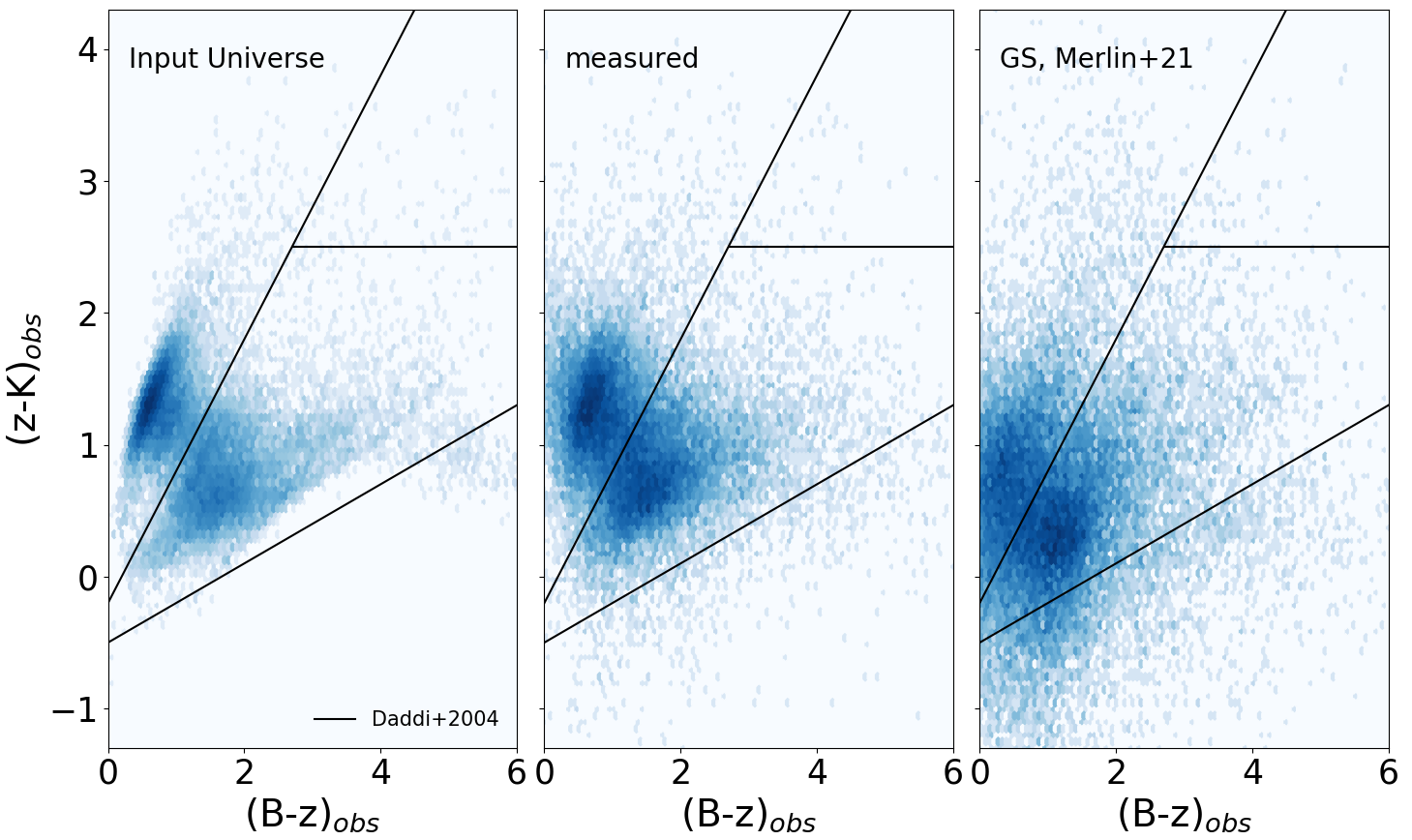}
        \caption{\label{bzk} $BzK$ diagrams for (left to right) (i) the IU fluxes of the detected sources, (ii) the measured colors of the same sources (matched with the IU), (iii) a sample of CANDELS sources from the deep region of the GS mosaic, with photometry from \cite{merlin21}.}
    \end{figure}

    We plot the $BzK$ colors of the simulated objects in Fig. \ref{bzk}, where the true colors (i.e., the ones obtained using IU fluxes) and the colors for a sample of the deep GS mosaic from \cite{merlin21} are also shown. The overall arrangement of the three distributions is indeed consistent, and we checked that star-forming sources at $1.4 < z_{true} < 2.5$ are reasonably well isolated in the upper left region of the diagram. 

    \subsection{Estimated physical properties}

    We finally checked the accuracy in the estimates of the redshift and the stellar mass of the detected galaxies.\\   
    To this aim, we performed a SED-fitting procedure with the code \textsc{zphot} \citep{font00}, adopted in many studies (e.g., \citealt{castellano16,santini15,merlin21}). We use a library of template galaxy SEDs identical to the one used in \cite{merlin21}. 
    
    \subsubsection{Photometric redshift}

\begin{figure}[h]
        \centering
        \resizebox{\hsize}{!}{\includegraphics{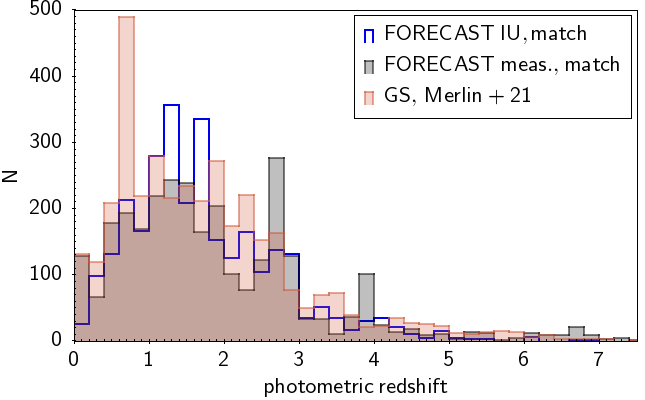}}
        \caption{Results of the SED-fitting photometric redshifts estimate using \textsc{zphot}; the plot shows the distribution of the redshifts of the sources with a IU match, both estimated with \textsc{zphot} (black shade) and from the input catalog (blue line); also shown is the distribution of a sample of sources from GOODS-South DEEP, taken from the ASTRODEEP catalog (\citealt{merlin21}, yellow shade). The histograms are normalized to allow for easier comparison.}
        \label{zdistr} 
    \end{figure}

\begin{figure}
     \centering
    \begin{subfigure}[t]{0.35\textwidth}
        \raisebox{-\height}{\includegraphics[width=\textwidth]{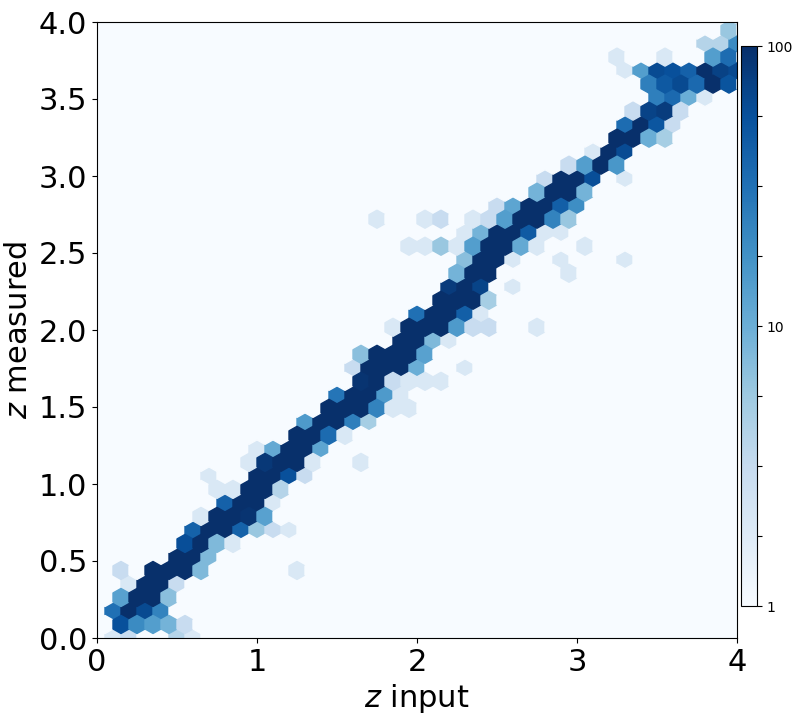}}
    \end{subfigure}
    \begin{subfigure}[t]{0.35\textwidth}
        \raisebox{-\height}{\includegraphics[width=\textwidth]{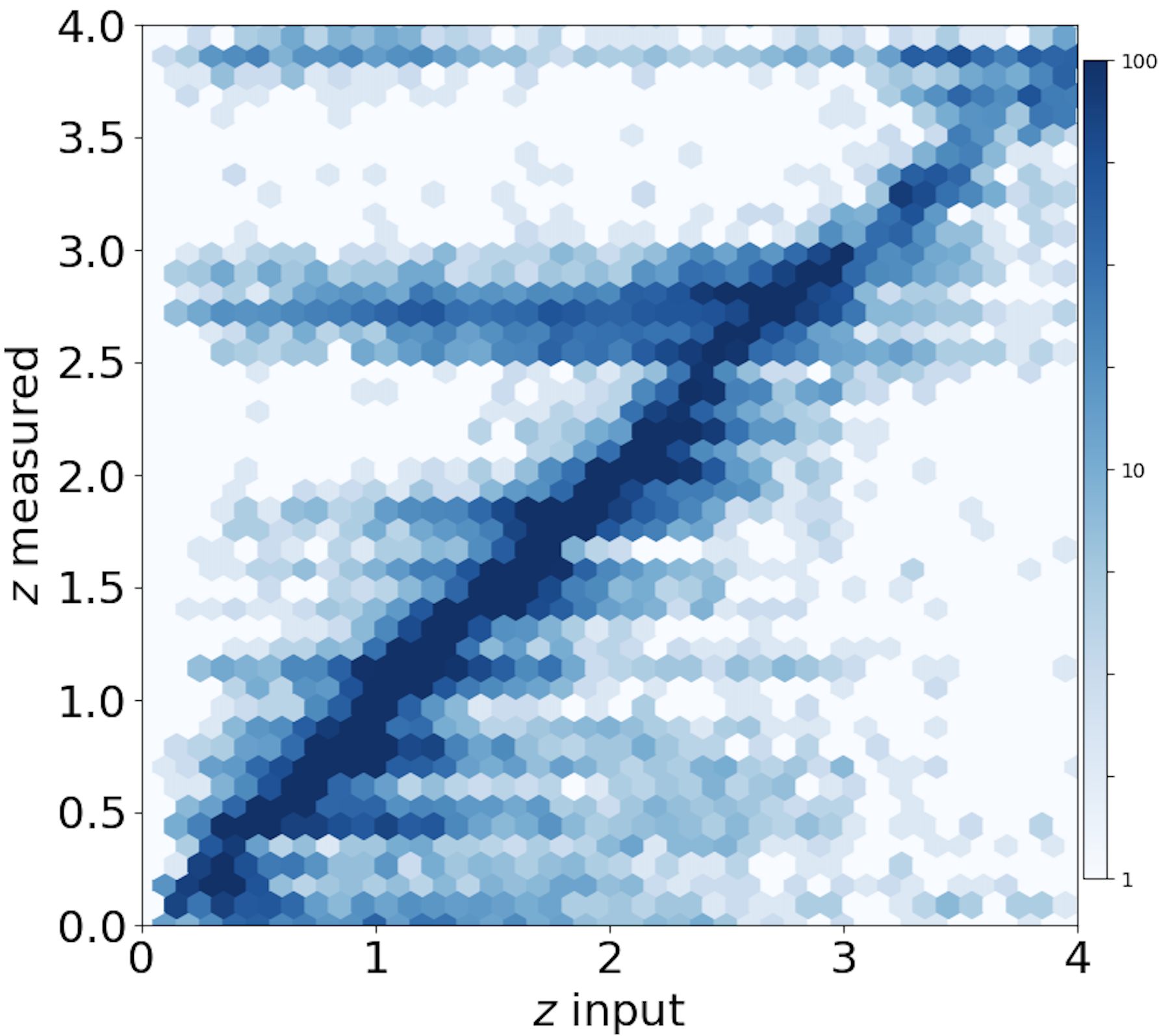}}
    \end{subfigure}
    \begin{subfigure}[t]{0.35\textwidth}
        \raisebox{-\height}{\includegraphics[width=\textwidth]{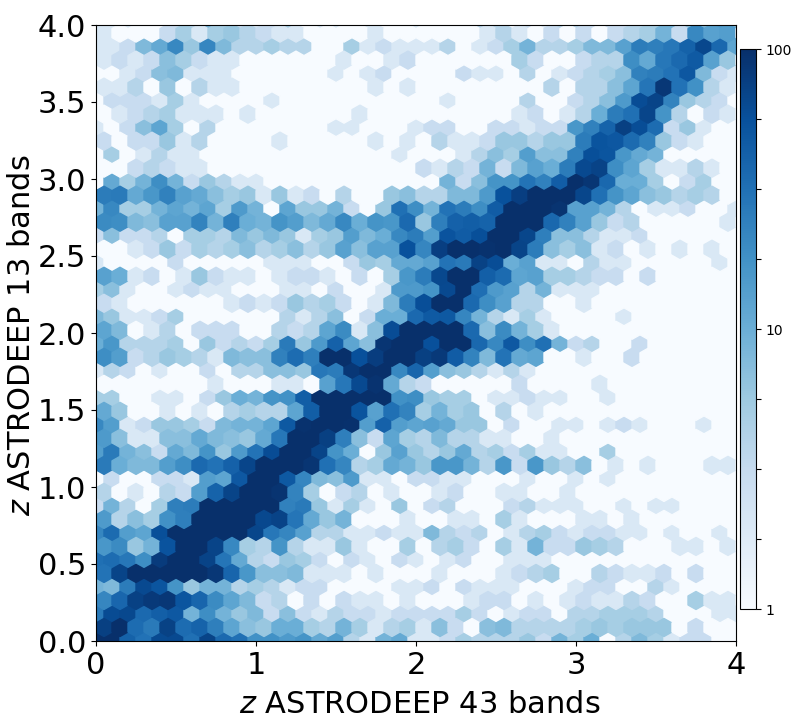}}
    \end{subfigure}
    \caption{
        Results of the SED-fitting photometric redshifts estimate using (from top to bottom): (i) \textsc{zphot} and the true fluxes of the input catalog; (ii) \textsc{zphot} and fluxes measured on the $H$ band simulated image; (iii) \textsc{zphot} on GOODS-South ASTRODEEP data.
        The plot shows the estimated redshifts versus the true redshifts from the input catalog (top and central panel); the redshifts estimated with the 13 bands used in the simulations versus the ones in the ASTRODEEP 43 bands catalog (\citealt{merlin21}, bottom panel).}
       \label{photoz} 
    \end{figure}

    The distribution of the measured redshifts is shown in Fig. \ref{zdistr}, together with the ones from the IU and from ASTRODEEP.\\
    We first checked that the library of models is sufficiently accurate, by estimating the redshifts using the true fluxes of the sources, while keeping the error budget of each source equal to the measured one. The result is in the top panel of Fig. \ref{photoz}. The agreement with the input redshifts is almost perfect, with a mean $dz=(z_{meas}-z_{true})/(1.0+z_{true})=-0.012 \pm 0.022$ for objects with $|dz| \leq 0.15$; the fraction of outliers (that have $|dz| > 0.15$) is very low ($\eta=0.21\%$).
      
    The distribution of the redshifts estimated using the measured fluxes is shown in the central panel of Fig. \ref{photoz}. 
    In general, the estimate seems to be reasonably accurate. 
    We note that the horizontal strips of catastrophic outliers are a typical feature of SED-fitting procedures, caused by a wrong interpretation of galaxy colors by the fitting algorithm, which interprets the red colors of a high redshift source as due to dust-obscured star formation in a low redshift object. The bulk of the objects are well recovered, with a mean $dz=-0.011 \pm 0.055$; however, the fraction of outliers is high ($\eta=25.2\%$). It suggests that most of the uncertainties and errors are caused by the scatter introduced by the photometric estimates rather than simply by the small number of bands. Of course, a larger number of bands helps minimizing the uncertainties in the fit.
    To further check this issue, we made a final test on the ASTRODEEP catalogs, but using only the same 13 bands simulated for this work to estimate photometric redshift. The results, performed using the same libraries of templates, are in the bottom panel of Fig. \ref{photoz}; however, it must be kept in mind that now we do not have a "true" value of the input redshift, but only the best estimates from spectroscopic or photometric data. While the results are better than the ones for the simulation ($dz=-0.003 \pm 0.055$), there are still a large amount of scatter and outliers ($\eta=20.8\%$), caused by the limited number of bands.
    
    \subsubsection{Galaxy stellar mass}

\begin{figure}
        \centering
        \begin{subfigure}[t]{0.4\textwidth}
            \raisebox{-\height}{\includegraphics[width=\textwidth]{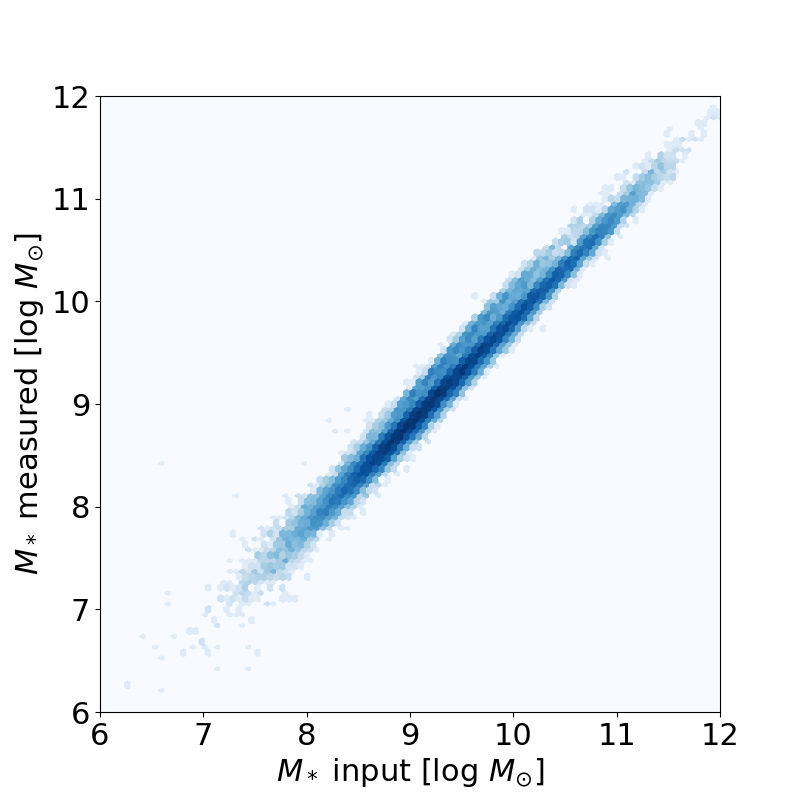}}
        \end{subfigure}

        \begin{subfigure}[t]{0.40\textwidth}
            \raisebox{-\height}{\includegraphics[width=\textwidth]{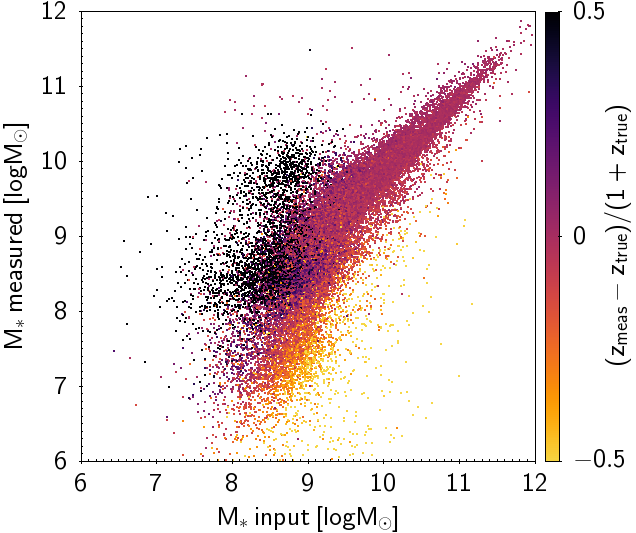}}
        \end{subfigure}
        \caption{
        Results of the SED-fitting stellar mass estimate using (from top to bottom): (i) \textsc{zphot} and both the true fluxes and the true redshifts; (ii) \textsc{zphot} and measured fluxes and redshifts, color-coded by error in redshift estimate, $(z_{meas}-z_{true})/z_{true}$. The plot shows the estimated stellar mass versus the true one from the input catalog.}
       \label{massz} 
    \end{figure}

    The top panel in Fig. \ref{massz} shows the comparison between the stellar masses estimated with \textsc{zphot}, fixing the redshift to the IU values and using the IU fluxes, and the true masses from the input catalog (defined as the sum of the masses of all the SSPs belonging to a given subhalo). Again, in this case the estimated values are in tight agreement with the true values in the mass estimates.
    We then replicate the same plot, but using measured redshifts and fluxes to estimate the masses of the sources with \textsc{zphot} (bottom panel of Fig. \ref{massz}). 
    Here the points are color-coded by the error the estimated redshift. The agreement is still good in general, although the scatter is quite large toward fainter masses.
    We note two interesting populations of sources: a group having masses underestimated by one order of magnitude, in the range $10^8 < M_{*,true} < 10^9$, and another group with masses overestimated by one order of magnitude ($10^8 < M_{*,true} < 10^9$).
    We see that the vast majority of the sources with underestimated masses also have underestimated redshifts (often by a factor $(z_{meas}-z_{true})/z_{true} \leq -0.5$): being considered as closer to the observer than they really are, they must be fitted with a low stellar mass to match the measured fluxes. A specular line of reasoning can be applied to objects with overestimated masses. We also note a group of sources with overestimated redshift which has masses correctly estimated (the dark points lying on the bisector of the distribution). We found that this feature is due to the underestimation of the $H$ flux for faint sources (see Sect. \ref{sec:mbp}), which causes their distance to be overestimated. 
    
    We also checked that estimating the masses using the measured fluxes but the IU true redshifts a less evident but non-negligible scatter is still present. We conclude that it is to be attributed to the uncertainties introduced by the photometric measurements, which will deserve further analysis in future work.
    
    \section{Code and data release}\label{sec:release}
    
    The FORECAST code (see Appendix \ref{code} for further details) is available to the community on our website \footnote{\url{http://www.astrodeep.eu/FORECAST}}, together with the CANDELS-like dataset analyzed in this work and a simulated dataset emulating $JWST$ observations, which is described in this section.

    \subsection{Data release of a JWST-like survey}\label{jwst}
    
    Together with the CANDELS mock observations, we have also produced and made public a second dataset, this time emulating $JWST$ observations of the same FoV we presented in the previous section. The set of data consists of a galaxy catalog and ten astronomical images. 
    To construct this dataset, we extended the light cone realized for the CANDELS emulation up to $z_s=20$, again exploiting the \textsc{IllustrisTNG} simulation. 
    The pixel scale is 0.031 arcsec (the typical value of JWST short wavelength detectors), yielding images on a grid of $\sim$ 750 million pixels. 
    We emulated ten $JWST$ bands: eight from NIRCam (F090W, F115W, F150W, F200W, F277W, F356W, F410M, F444W), and two from MIRI (F560W, F770W). For the post-processing we used the PSF models provided by STScI in the WebbPSF webpage\footnote{\url{https://jwst-docs.stsci.edu/jwst-near-infrared-camera/nircam-predicted-performance/nircam-point-spread-functions}}. 
    For this emulation we also decided to change the synthetic stellar population model by adopting \cite{gutkin16}, which includes rest-frame ultraviolet and optical nebular emission from \textsc{Hii} regions in star-forming galaxies in a wide range of chemical compositions.\\    
    \indent The image in the \textit{JWST}/NIRCam F090W filter is simulated with resolution and limit magnitude adopted by \cite{merlin22} for the \textit{JWST} GLASS survey \citep{treu22}. Additionally, the remaining seven broad-band images in NIRCam filters emulate the \textit{JWST} CEERS survey \citep{finkelstein22a}; we created two images in the CEERS \textit{JWST}/MIRI bands with resolution and depth adopted by \cite{papovich22}. We did not perform any analysis on this dataset, leaving it to future work.
    All the released data products are available on our website\footnote{\url{www.astrodeep.eu/FORECAST}}.
    
    \begin{table}[ht!]
        \caption{Summary of the instrumental PSF and depths adopted for the image simulations in the $JWST$ filters from \cite{merlin22, finkelstein22a, papovich22}.}
        \label{filterjwst}
        \centering
        \begin{threeparttable}
        \begin{tabular}[t]{cccc}
            \hline\hline
            Instrument&Filter&PSF(arcsec)&5$\sigma$ depth AB\\
            \hline
            NIR-Cam&F090W&0.035&28.78 \tablefootmark{a}\\
            &F115W&0.066&29.20\tablefootmark{b}\\
            &F150W&0.070&29.00\tablefootmark{b}\\
            &F200W&0.077&29.20\tablefootmark{b}\\
            &F277W&0.123&29.20\tablefootmark{b}\\
            &F356W&0.142&29.20\tablefootmark{b}\\
            &F410M&0.155&28.40\tablefootmark{b}\\
            &F444W&0.161&28.60\tablefootmark{b}\\
            MIRI &F560W&0.240&26.50\tablefootmark{c}\\
            &F770W&0.280&27.10\tablefootmark{c}\\
            \hline
        \end{tabular}
        \end{threeparttable}
        \tablefoot{
            \tablefoottext{a} {The limiting magnitude is measured in a 0.1" radius aperture, from \cite{merlin22}.\\}
            \tablefoottext{b} {The limiting magnitudes are measured in a 0.1" radius aperture, from \cite{finkelstein22a}.\\}
            \tablefoottext{c} {These depths are derived from tests on the MIRI images. They are measured in 0.45" diameter apertures (see \citealt{papovich22} for further details).}}
    \end{table}
    
    \section{Summary and conclusions}\label{sec:final}

    We have presented FORECAST, a new software package that performs forward modeling of the output of cosmological hydrodynamical simulations to create realistic synthetic astronomical images. Starting from the physical properties of the simulated stellar resolution elements provided in the output snapshots of a hydrodynamical simulation, the software computes their expected fluxes, accounting for $k$ correction, attenuation by dust and by the intergalactic medium, and arranges them to produce images to which background noise, PSF smoothing and potentially other observational features can be added. The simulated galaxies are built particle by particle, and therefore they do not have smooth, analytical light profiles; instead, they have realistic morphologies and fluxes, computed from their complex star formation histories. The simulated images can be processed and analyzed with the same methods and tools used in real data analysis, and directly compared in a fully consistent way to the results from real observational data. FORECAST is a flexible tool that can produce realistic images, enabling the analysis of possible systematics and biases arising in observations due to image processing, the choice and limitations of the algorithms used to detect, deblend, and measure galaxy fluxes, as well as the physical assumptions in a SED-fitting procedure.\\

    To test FORECAST, we built a light cone between $z=0$ and $z=7$, emulating the GOODS-South CANDELS field, creating scientific images in eight $HST$ bands (ACS: $B$435, $V$606, $I$814, and $Z$850; and WFC3: $Y$105, $J$125, $JH$140, and $H$160), one $VLT$ band (HAWK-I $K_s$), and four Spitzer bands (IRAC CH1, CH2, CH3, CH4), considering the deep region of the field as a reference for the synthetic image depths.
    
    The simulated field of view has an area of 200 sq. arcmin, and the fluxes are mapped on a grid of 200 million of pixels, resulting in a pixel scale of 0.06 arcsec, a typical resolution of real $HST$ bands (we created all bands with the same pixel scale, avoiding rebinning procedures).
    This light cone includes a great diversity of galaxies over a large range of mass and star formation rates, metallicities, ages, star formation histories generated by the complex interplay of the diverse astrophysical processes (cooling, star formation, feedback, and dynamical evolution and interactions). The final products, the images in the 13 bands, are generated adding noise and PSF to the outputs of FORECAST with our post-processing procedure. We found that the simulated images offer a realistic representation of many observational features; we verified this using standard techniques used for the photometric analysis of real images.\\

    We performed the detection on the simulated $H$ band using \textsc{SExtractor}, and we then measured the fluxes of the detected sources using aperture photometry with \textsc{a-phot} on the 13 simulated images. The flux of the sources is generally well recovered in all bands, with a slight underestimation at faint magnitudes due to the measurements on the $H$ band, which is used to derive the fluxes in the remaining bands.\\
    
    We then checked the number counts of the simulated sources, comparing the counts as a function of the $H$ magnitude between a sample of simulated objects (with fluxes taken both from the Input Universe and from the detection process), and a sample of objects in the CANDELS GOODS-South area. We found that the number of objects detected on the simulated image is consistent with the Input Universe (IU) up to mag$H\sim25$, after which their counts begin to be less than expected, also compared with the trend in CANDELS. We determined that there are two contributing factors to the discrepancy between the counts in the range $26<H<27$: firstly, the \textsc{IllustrisTNG} IU seems less populated than expected; secondly, there is a large fraction of galaxies that are either blended with larger objects or too faint to be detected due to their fragmented morphologies. 
    However, the identification of significant overdensities in the GOODS-South field, spanning redshifts z=0.6-3.7  \citep{castellano07,salimbeni09,kurk09,kang09,castellano11}, suggests that the galaxy counts in the CANDELS catalog might have been impacted by this clustering, potentially resulting in an increase in the galaxy counts.\\
      
    We estimated the physical properties of the galaxies detected on the simulated images via SED-fitting. While the redshifts are perfectly recovered in ideal conditions (i.e., using true IU fluxes), a noticeable amount of scatter is introduced using the measured fluxes. 
    The accuracy is almost perfect also in the estimate of the stellar masses if the photometry is ideal (i.e., fitting the true fluxes at the true redshifts). 
    However, a mild scatter emerges if the measured fluxes are fitted at the measured redshifts, mostly caused by the error committed in measuring the $H$ flux at faint magnitude, which is spread in the other bands; for a subsample of sources, an error of one order of magnitude in the estimate mainly depends on the propagation of the error on the photo-$z$ estimate.\\   
    
    We want to remark that the realization of these synthetic images is the first attempt of forward modeling as much physics as possible from hydrodynamical simulations, and the tests performed in this work must be intended as a first quality check.\\
    
    Future work will include (i) implementing additional effects in the light cone: adding Milky Way stars and local objects, Active Galactic Nuclei, the absorption due to Milky Way gas and dust, the effect of lensing; (ii) implementing options to allow for more flexibility, e.g., giving the user the possibility to choose a preferential position to extract the light cone.\\

    We make the simulated CANDELS dataset publicly available. We also release a set of images simulated in ten JWST bands, and the corresponding Input Universe catalog containing simulated physical properties and simulated true fluxes of the galaxies.    
    \newline
    \indent As new upcoming observational instruments will allow us to probe the Universe to an unexplored extent, numerical tools like FORECAST will help us to capture the significance of their exploration, improving the synergism between observations and theory.
    The next few years will revolutionize our understanding of the Cosmos and will make us more aware about the Universe we inhabit. 

    \begin{acknowledgements}
    
    The IllustrisTNG simulations were undertaken with computational time awarded by the Gauss Centre for Supercomputing (GCS) under GCS Large-Scale Projects GCS-ILLU and GCS-DWAR on the GCS share of the supercomputer Hazel Hen at the High-Performance Computing Center Stuttgart (HLRS), as well as on the machines of the Max Planck Computing and Data Facility (MPCDF) in Garching, Germany.\\
    We thank the CINECA award under the ISCRA initiative, for the availability of high-performance computing resources and support. We thank the INAF computing system PLEIADI, for the availability of high-performance computing resources and support.\\
    Carlo Giocoli acknowledges support from the PRIN-MIUR 2017 WSCC32 ZOOMING, the ASI n.2018-23-HH.0, the INAF grant under the "Bando PrIN 2019", PI: Viola Allevato, the INAF theory grant 2022: Illuminating Dark Matter using Weak Lensing by Cluster Satellites, PI: Carlo Giocoli.
    \end{acknowledgements} 

\bibliographystyle{aa}

    \begin{appendix}

    \section{Mock light cone with IllustrisTNG100}
    
    In order to convert the output of the \textsc{IllustrisTNG100-1} simulation into mock astronomical images, we extracted the physical properties of stellar particles and gas cells required by FORECAST, as described in Sects. \ref{light} and \ref{dustatt}.
    In particular, for stellar particles ("\textsc{PartType4}") we extracted "\textsf{Coordinates}";  "\textsf{Masses}"; "\textsf{GFM$\_$InitialMass}"; "\textsf{GFM$\_$Metallicity}"; "\textsf{GFM$\_$StellarFormationTime}", the latter to compute the ages of the SSPs. FORECAST also requires information about the subhalo membership of the particles (as it is defined by the simulation procedure), but for TNG it is not stored in the snapshots for stellar particles and gas cells, while subhalos in subhalo catalogs have information about their particle and cell members. Thus, we reconstructed the membership of stellar particles and gas cell in reverse, with a designed algorithm that uses \textsc{SubFind} subhalo catalogs and \textsf{Offset} files, exploiting the specific organization of halos and subhalos within the catalog files.      
    
    Concerning gas cells ("\textsc{PartType0}"), we extracted their "\textsf{Coordinates}"; "\textsf{Masses}"; "\textsf{Density}"; "\textsf{GFM$\_$Metallicity}"; "\textsf{ElectronAbundance}", which is gas cell fractional electron number density $x_e$, with respect to the total hydrogen number density, that is $n_e=x_e n_H$, and "\textsf{InternalEnergy}", which is gas cell internal (thermal) energy per unit mass $u$, both needed to derive the neutral hydrogen column density $N_{HI}$ (see Eq. \ref{tau}). The IDs of gas cells are exploited to derive their subhalo membership, as done for star particles. \footnote{The available fields for stellar particles and gas cells, their units, and descriptions are available at \url{https://www.tng-project.org/data/docs/specifications/}.}
    
    The neutral hydrogen column density within each gas cell is not available in the simulation output, so we estimated it as follows, taking advantage of already accessible properties. 
    The fraction of neutral hydrogen with respect to the total hydrogen number density within each gas cell, "\textsc{ NeutralHydrogenAbundance}" alias $x_{HI}$, necessary to compute the neutral hydrogen column density, is available in the \textsc{IllustrisTNG} output only at certain snapshots, the so-called "full snapshots"; the remaining "mini snapshots" only have a subset of particle fields available, and they do not include $x_{HI}$. To overcome this shortage, we perform a fourth-degree least squares polynomial fit on the relation between the neutral hydrogen-to-gas mass ratio $M_{HI}/M_g$ and the temperature of the gas $T_g$ in the "full snapshots"

\begin{figure}[h!]
        \centering
        \includegraphics[width=0.48\textwidth]{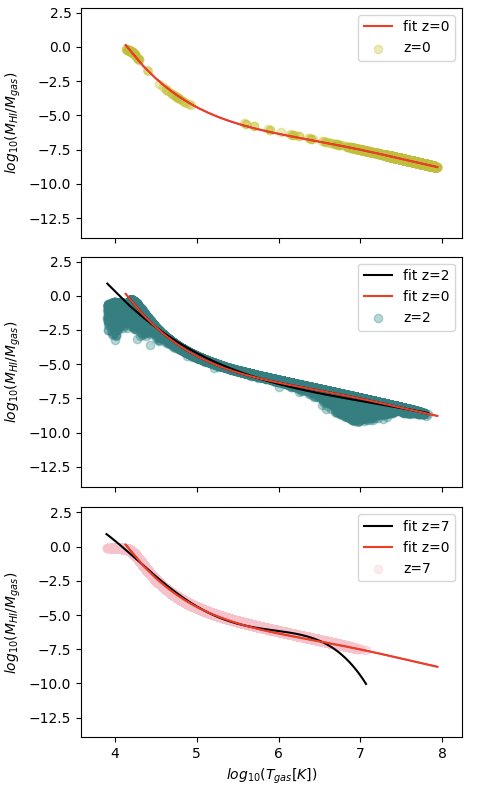}
        \caption{Scaling relation between neutral hydrogen-to-gas mass ratio $M_{HI}/M_g$ and the temperature of the gas $T_g$ in the "full snapshots" of the \textsc{IllustrisTNG}100 simulation, where the neutral hydrogen fraction is available ($z=0,2,7$). The red curve is the best fourth-grade polynomial fit at $z=0$; black curves are best fits for the data points (dots) at the redshift of each panel.}
        \label{fit}
    \end{figure}

    When performing the polynomial fit to the distribution of gas cells' values in the "full snapshots", we find that the red curve in Fig. \ref{fit}, representing the best fit at $z=0$, fits the relation up to $z=7$ with sufficient accuracy, so we decide to use the polynomial coefficients of the best fit at $z=0$ to derive $M_{HI}/M_g$ throughout the full light cone (in each snapshot). To maintain the fitting curve within a valid physical range, we set to 0 the values of $log(M_{HI}/M_g)$ that exceed 0. The best fit coefficients of the relation between $M_{HI}/M_g$ and $T_g$ are given in Table \ref{coeff}.

    \begin{table*}[ht!] 
        \centering
        \caption{Fourth degree polynomial fit coefficients at $z=0$.}
        \label{coeff}
        \begin{tabular}{ cccccc }
        \hline\hline
         \multicolumn{6}{c}{$ \frac{M_{HI}}{M_g}  = a_1 \, T^4_g + a_2 \, T^3_g + a_3 \, T^2_g + a_4 \, T_g + a_5$} \\
        \hline
        z&$a_1$&$a_2$&$a_3$&$a_4$&$a_5$\\
        0 & 2.31105e-02 & -8.91893e-01 & 1.14885e+01 & -6.27073e+01 & 1.18956e+02\\
        \hline
        \end{tabular}
    \end{table*}

    Gas number density of each gas cell is derived as $n_g = \rho_g / m_p$, where $m_p$ is the proton mass. Therefore, it is possible to compute the neutral hydrogen number density $n_{HI}=x_{HI} \, x_H \, n_g$, from which we derive the neutral hydrogen column density of each gas cell, assuming their volume as cubic with side $L_g$ 
    
    \begin{equation}
        N_{HI} = n_{HI} \cdot L_g.
    \end{equation}
    
    \section{FORECAST code}\label{code}   
    FORECAST is a robust and flexible code. Its main body, written in C and C++, is supported by independent libraries to make the code more readable and user-friendly. It requires the following C/C++ standard libraries: gsl\footnote{\url{https://www.gnu.org/software/gsl/}}, openBLAS\footnote{\url{https://www.openblas.net/}}, LAPACK\footnote{\url{http://www.netlib.org/lapack/}}, CCfits\footnote{\url{https://heasarc.gsfc.nasa.gov/fitsio/CCfits/}},  CFITSIO\footnote{\url{https://heasarc.gsfc.nasa.gov/fitsio/}}, FFTW \footnote{\url{https://www.fftw.org/}}, Eigen\footnote{\url{https://eigen.tuxfamily.org/}},  Armadillo\footnote{\url{http://arma.sourceforge.net/}}, H5Cpp\footnote{\url{http://h5cpp.org/}}, HDF5 C++\footnote{\url{https://www.hdfgroup.org/}}, and a gcc compiler.\\   
    \indent In the input configuration file of the code, the user chooses the image simulation parameters (e.g., the dimension of the FoV, the filters, the hydrodynamical simulation; see Sect. \ref{inpar}).\\   
    \indent The code requires the input files of the chosen hydrodynamical simulation to build the light cone and produce the final images. The data products of the numerous currently available simulations are organized differently, changing from one to another simulation, and are stored with different formats; as example \textsc{IllustrisTNG} stores a single snapshot in multiple \texttt{.hdf5} files, while in the \textsc{eagle} simulation \citep{eagle15} the snapshots are available for public download via an SQL web interface. Thus, the code requires these data products to be uniformed in a specific format in order to be easily read and processed.\\   
    \indent The code pipeline consists of four modules. The architecture of each module is not intrinsically parallel (e.g., it does not exploit MPI protocols), but it has been designed to allow the user to independently run it on multiple snapshots to realize multiple light-cone partitions simultaneously.\\   
    \indent We release a beta version of the code, which can be read and improved by the scientific community with a request for access to its source through our website\footnote{\url{www.astrodeep.eu/FORECAST}}. 
    
    \subsubsection{Pipeline}
    
    The FORECAST code, currently available in beta version, is structured into four interconnected modules, where each module relies on the output of the previous one. The first and third modules require as input file the snapshots of a hydrodynamical simulation to extract relevant properties of the simulated stellar and gas resolution elements. Currently, the code reads input files in the \textsc{IllustrisTNG} format. Users are required to convert their data into the TNG format (\texttt{.hdf5} files with the same column names as TNG columns) to ensure the effective utilization of the code. Future updates to the code will include additional scripts to read input files from multiple hydrodynamical simulations in different formats, such as the EAGLE Project \citep{sch15} and the Simba Simulation \citep{simba19} data products.
    
    \indent The first module handles the construction of the light cone, exploiting the data products of a chosen hydrodynamical simulation (see Sect. \ref{lc}). The resulting output is an \texttt{ASCII} file containing the properties of the SSP elements within the field of view, including their IDs, coordinates, redshift, and physical characteristics. This file serves as the input for the subsequent module. This step might be skipped if a user already has their light cone, as long as the input file for the next module is written in the proper format.\\ 
    The second module computes the dust-free flux of each SSPs included in the FoV (see Sect. \ref{light}). It assembles an \texttt{ASCII} file with the properties of the star particles and their dust-free fluxes in chosen filters.\\
    The third module addresses the computation of dust-corrected fluxes and it requires the data products of the hydrodynamical simulation to extract the properties of gas resolution elements belonging to the sources included in the FoV. In output it is given the same file produced by the previous module, also including dust-corrected fluxes for stellar particles, and the gas mass-weighted mean of the gas metallicity and the neutral hydrogen column density. The final module adds the IGM correction to dust-corrected fluxes, producing the final output catalog, which includes the physical properties of the stellar particles and their corrected fluxes, and the mean properties of the gas.\\   
    \indent An independent C++ script handles the arrangement of the fluxes on a grid of pixels with $N_{pix}$-per-side chosen by the user.\\   
    Two additional independent scripts, written in \textsf{python}, are made available (i) to build the galaxy catalog, in \texttt{ASCII} format, from the particle catalog given in output by FORECAST (see Sect. \ref{app:output} for a full description of the output); (ii) to post-process the FORECAST images with our noise and PSF pipeline (see Sect. \ref{sec:realism}).
   
    \subsubsection{Output}\label{app:output}
    The output of the code is the catalog including the physical properties and the true fluxes of the simulated stellar particles. It is used to build the galaxy catalog (see Appendix \ref{app:catalog} for a full description of the fields included in the galaxy catalog).
    The catalogs, both the particles and the galaxy ones, have different sizes depending on the number of particles (or galaxies) included in the FoV, and typically grow in size as the redshift increases since more structures are included. The total size of output files is $\sim$ 72 GB.
    The output images are recorded on 16-bit floating-point \texttt{.fits} files. Each plane (projection on a bi-dimensional map of fluxes from a volume of the Universe included in the field of view, in a redshift range) occupies $\sim$ 3 GB, while the size of the final stacked image is $\sim$ 5 GB.
    
    \subsubsection{Memory consumption and running time}  
    We performed some tests and realized the light cone and the images presented in this work on the GALILEO100 supercomputer located at CINECA \footnote{The description of the GALILEO100 architecture is available at \url{https://wiki.u-gov.it/confluence/display/SCAIUS/HPC+User+Guide}.}.
    The peak of memory consumption is reached during the post-processing of the dust, and in particular, during the intensive manipulation of gas elements: in each run corresponding to a snapshot, hundreds of millions of gas components are tracked in front of millions of stellar particles, and some gas properties might have to be computed and assigned (e.g.,  neutral hydrogen column density, see Sect. \ref{dustatt} for more details). The typical memory consumed with these procedures in the third module is currently 180 GB, which is almost five times the amount of memory requested in the second module, during the convolution and integration of the SSPs SED within filters, which are memory-intensive operations (40 GB RAM) performed with well-optimized routines.\\ 
    \indent The code architecture, which has not been conceived as parallel, can be improved by implementing an MPI protocol to minimize memory consumption.\\
    \indent The building of the light cone structure, the extraction of stellar particles properties from snapshot files, the arrangement of stellar particles in the observing cone following geometrical cuts, and the selection of the stellar particles in the FoV (first module) take $\sim 2$ to 15 minutes, depending on the number of stellar particles included in the simulation volume. This is true also for the calculations related to the IGM absorption in the fourth module. In the second module, the connection of each SSP included in the FoV with the corresponding synthetic stellar population model is computationally inexpensive, while the conversion of the rest-frame spectrum into the observer-frame flux per unit wavelength, and its convolution and integration with the filter response are time-consuming operations, by order of $\sim 15 $ ms per stellar particles, translating into maximum $\sim 0.6$ day of running per single snapshot if the particle budget is high ($\sim 1.8 - 3.5$ million of stellar particles in the FoV). Concerning the implementation of dust effects (third module), the tracking of gas elements in front of star particles and the manipulation of their properties to derive the relevant quantities used to turn dust-free in dust-corrected magnitudes (e.g.,  $N_{HI}$) is slower ($\sim1$ day per snapshot).

    \section{The Input Universe catalog} \label{app:catalog}
    
    The Input Universe catalog is a file that collects all the information on the sources included in the simulated FoV, before they are post-processed and measured, that is their true values. It is built from the particle catalogs, which are the output of the code at each run (see Appendix \ref{code}). The available fields, with their units and their description, are listed in Table \ref{galaxycat}.

    \begin{table*}[h!]        
        \centering
        \caption{Galaxy catalog content - Input Universe}
        \label{galaxycat}
        \begin{tabular}{p{0.15\textwidth}p{0.08\textwidth}p{0.62\textwidth}}
            \hline\hline
            Field&Units&Description\\
            \hline
            subhaloID&-&identifier of Subhalo at that redshift; not unique; it could repeat throughout the full catalog.\\
            redshift&-&redshift of Subhalo.\\
            stellar mass&$M_{\odot}$&stellar mass of Subhalo.\\
            gas mass&$M_{\odot}$&gas mass of Subhalo.\\
            DM mass&$M_{\odot}$&DM mass of Subhalo.\\
            SFR&$M_{\odot}yr^{-1}$&instantaneous star formation rate of Subhalo.\\
            metallicity&$Z_{\odot}$&stellar mass-weighted metallicity of Subhalo.\\
            age&Gyr&stellar mass-weighted age of Subhalo.\\
            oldest\,\, SSP age&Gyr&age of the oldest SSP of Subhalo.\\
            $(x_c, y_c)$&pix&coordinates of the center of Subhalo.\\
            $R_{max}$&pix&maximum radius of Subhalo; it matches the further SSP from the center.\\
            $N_{SSP}$&-&number of SSP in Subhalo.\\
            $F_i$&$\mu$J&observer-frame, integrated flux in filter $i$.\\
            $F_{i,3pix}$&$\mu$J&observer-frame, integrated flux in filter $i$, in $R=3$ pix.\\
            $R_{i,hl}$&pix&half-light radius for $F_{i,R_{i,hl}}$.\\
            $F_{i, R_{i,hl}}$&$\mu$J&observer-frame, integrated flux in filter $i$, in $R_{i,hl}$.\\      
            \hline
        \end{tabular}
    \end{table*}
    
    \end{appendix}

\end{document}